\documentclass[iop,twocolumn,twocolappendix]{emulateapj}
\usepackage{natbib}
\usepackage{hyperref}
\usepackage{graphicx}
\usepackage{epsfig}
\usepackage{textcomp}
\usepackage{subfigure}
\usepackage{booktabs}
\usepackage{color}
\usepackage{soul}
\slugcomment{}
\shorttitle{Subtle variations of the 11 and 12.7 \mt PAH bands}
\shortauthors{M. J. Shannon et al.}

\begin{document}
\newcommand{\mt}{\textmu m~}
\newcommand{\m}{\textmu m}
\newcommand{\HII}{\mbox{H\,\textsc{ii}}}%
\newcommand{\updated}{\textcolor{magenta}{[Updated]}~}
\newcommand{\incomp}{\textcolor{red}{[Incomplete]}~}
\newcommand{\choop}{CH$_{\text{oop}}$~}

\title{Interpreting the subtle spectral variations of the\\ 11.2 and 12.7 \mt polycyclic aromatic hydrocarbon bands}

\author{M. J. Shannon\altaffilmark{$\dagger$,1}, D. J. Stock\altaffilmark{1}, E. Peeters\altaffilmark{1,2}}

\altaffiltext{$\dagger$}{Contact: mshann3@uwo.ca}
\altaffiltext{1}{Department of Physics and Astronomy, University of Western Ontario, London, ON, N6A 3K7, Canada}
\altaffiltext{2}{SETI Institute, 189 Bernardo Avenue, Suite 100, Mountain View, CA 94043, USA}

\begin{abstract}
We report new properties of the 11 and 12.7 \mt emission complexes of polycyclic aromatic hydrocarbons (PAHs) by applying a Gaussian-based decomposition technique. Using high-resolution \textit{Spitzer} Space Telescope data, we study in detail the spectral and spatial characteristics of the 11 and 12.7 \mt emission bands in maps of reflection nebulae NGC 7023 and NGC 2023 (North and South) and the star-forming region M17. Profile variations are observed in both the 11 and 12.7 \mt emission bands. We identify a neutral contribution to the traditional 11.0 \mt PAH band and a cationic contribution to the traditional 11.2 \mt band, the latter of which affects the PAH class of the 11.2 \mt emission in our sample. The peak variations of the 12.7 \mt complex are explained by the competition between two underlying blended components. The spatial distributions of these components link them to cations and neutrals. We conclude that the 12.7 \mt emission originates in both neutral and cationic PAHs, lending support to the use of 12.7/11.2 intensity ratio as a charge proxy.
\end{abstract}


\section{Introduction}

Prominent infrared (IR) emission bands between 3 and 20 \mt are observed in many astronomical environments. These spectral features are attributed to the vibrational fluorescence of polycyclic aromatic hydrocarbons (PAHs), which are electronically excited by ultraviolet (UV) photons. PAH molecules are composed of a hexagonal honeycomb carbon lattice, typically containing 50-100 carbon atoms, with a dusting of hydrogen atoms about the periphery. Compact structures are generally the most stable, but a variety of PAH shapes and sizes are expected to exist in space \citep{vanderzwet1985,allamandola1989,jochims1994}.

The strongest PAH emission bands are observed at 3.3, 6.2, 7.7, 8.6, 11.2, and 12.7 \m. A variety of weaker bands are also seen in the observational spectra (at, e.g., 11.0, 12.0 and 13.5, 14.2, 16.4 \m). The PAH bands can be associated with the following vibrations: C-H stretching (3.3 \m); C-C stretching (6.2 \m); C-C stretching and C-H in-plane bending (7.7, 8.6 \m); and C-H out-of-plane bending (hereafter CH$_{\text{oop}}$---PAH bands in the 10-15 \mt region). It is the number of adjacent C-H groups that determines the wavelength of the emission in the 10-15 \mt region, i.e.solo, duo, trio and quartet C-H groups.

The relative emission intensities in these bands are known to be highly variable between sources and within individual resolved objects (e.g. \citealt{hony2001,galliano2008b,stock2014,shannon2015}). The charge state of the PAH population has been identified in laboratory studies as the most important parameter in driving variations in the relative emission intensities, sometimes reaching one order of magnitude between charge states \citep{allamandola1999,galliano2008b}. Likewise, the profiles are known to vary in shape and peak position, which have been linked to object type (e.g. \citealt{peeters2002,vandiedenhoven2004}). The variability of the PAH profiles is thought to represent differences in PAH sub-populations, possibly in, e.g., size or structure (e.g. \citealt{hudgins2005,sloan2007,candian2012,sloan2014}; see \citealt{peeters2011} for a detailed overview).

Decomposing the PAH emission bands with a mixture of functions (e.g. Gaussians, Lorentzians, Drude profiles) is one way to investigate the origins of the observed spectral variability \citep{peeters2002,smith2007,galliano2008b,boersma2012}. A recent result by Peeters et al. (2015) showed that the 7.7 and 8.6 \mt emission bands can be decomposed into four Gaussian components, revealing that at least two PAH sub-populations contributed to this emission. Motivated by this result, we apply here a similar approach to the 11 \mt emission complex (i.e., both the 11.0 and 11.2 \mt bands) and the 12.7 \mt emission complex. Since the variations of the 11 and 12.7 \mt complexes are relatively minor when compared to the 7.7 and 8.6 \mt emission bands, it is critical to examine high spectral resolution observations. In addition, if band substructure indeed traces PAH sub-populations, the astronomical data considered must span a sufficiently wide swath of physical conditions, such that any intrinsic differences are reflected in the observational band profiles.

We present here new decompositions of the 11 and 12.7 \mt emission complexes in high-resolution \textit{Spitzer}/IRS maps of RNe and a star-forming region in order to understand the PAH sub-populations that produce the blended emission bands. We organize the paper as follows: the targets, observations, and continuum determination are presented in Section~\ref{sec:obs}. The spectral variability in the spectra prior to any further analysis are examined in Section~\ref{sec:specprof}. We introduce new methods for decomposing the 11 and 12.7 \mt PAH emission bands in Section~\ref{sec:methods}. Results are presented in Section~\ref{sec:results} and we discuss the implications of these results in Section~\ref{sec:discussion}. We present a brief summary in Section~\ref{sec:conclusion}.

\section{Observations and Data}
\label{sec:obs}

\subsection{Target selection and observations}

We chose targets with \textit{Spitzer}/IRS-SH maps that exhibit strong emission in the 11 and 12.7 \mt PAH complexes. The chosen sources were the RNe NGC 7023, NGC 2023 (two pointings---North and South), and the star-forming region M17.

\begin{figure*}
\begin{center}
    \subfigure{
	\includegraphics[height=6cm]{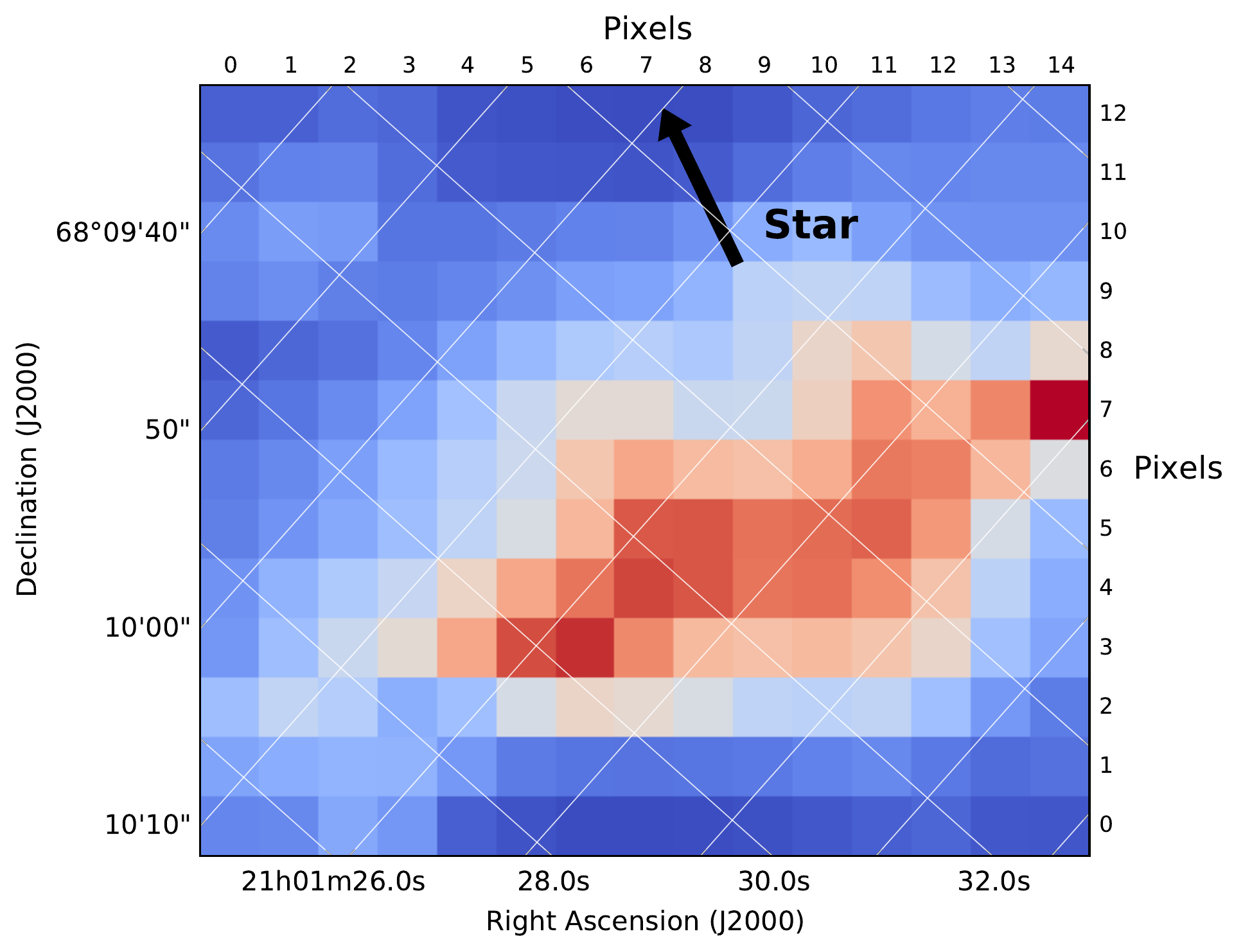}
	\label{fig:astro_ngc7023}
    }
    \subfigure{
	\includegraphics[height=6cm]{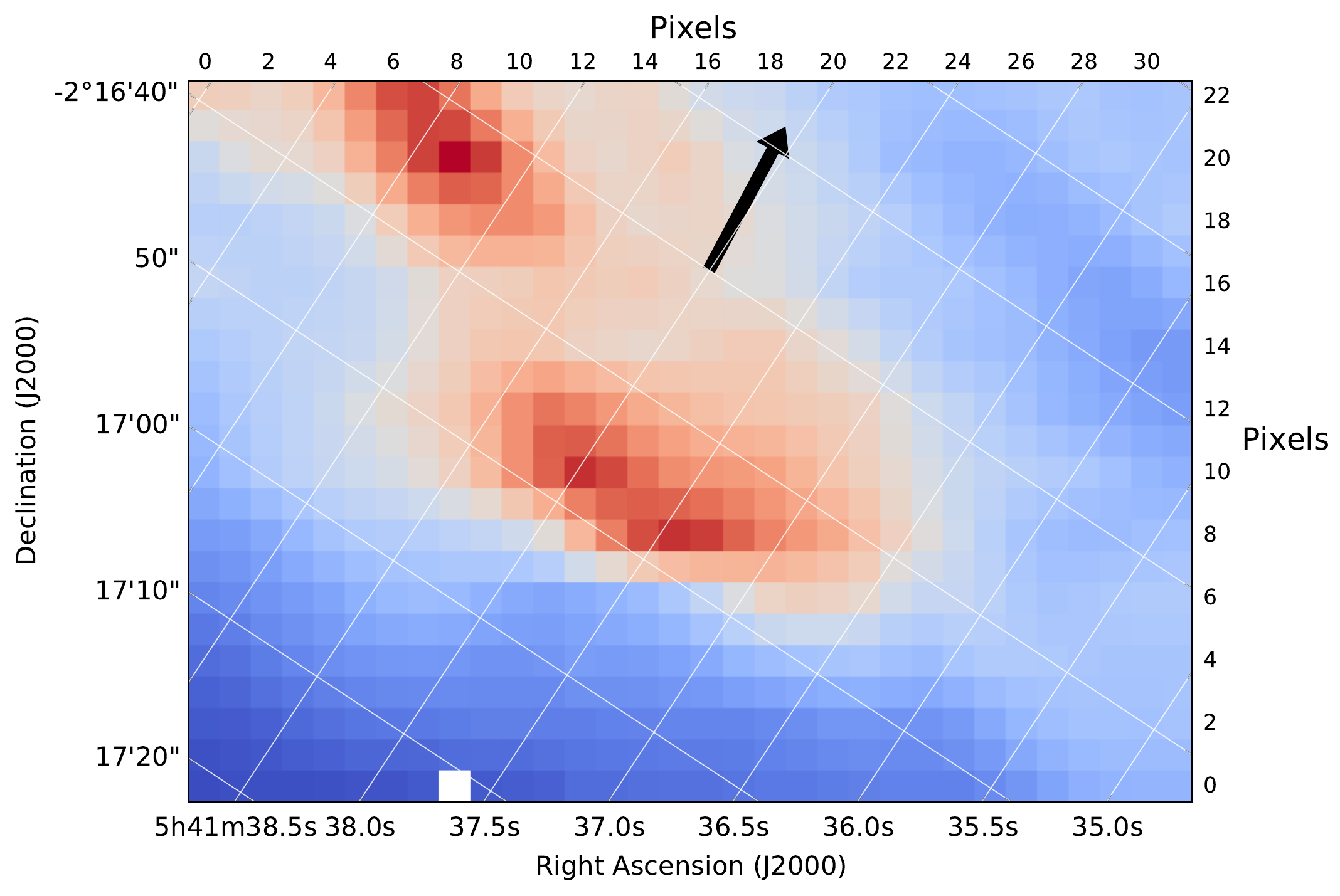}
	\label{fig:astro_ngc2023S}
    }
\caption{Astrometry for NGC 7023 (left) and NGC 2023 South (right). The normalized flux of the traditional 11.2 \mt band is shown in each map (cf. Figs.~\ref{fig:map_7023} and ~\ref{fig:map_2023S}). The illuminating star for NGC 7023 is HD 200775, 11\arcsec~beyond the boundary of the image along the indicated radial vector (black arrow). For NGC 2023 South, the illuminating star is HD 37903, and it is approximately 42\arcsec~from the boundary of the image.}
\label{fig:astrometry}
\end{center}
\end{figure*}

\begin{figure*}
\begin{center}
    \subfigure{
	\includegraphics[height=6cm]{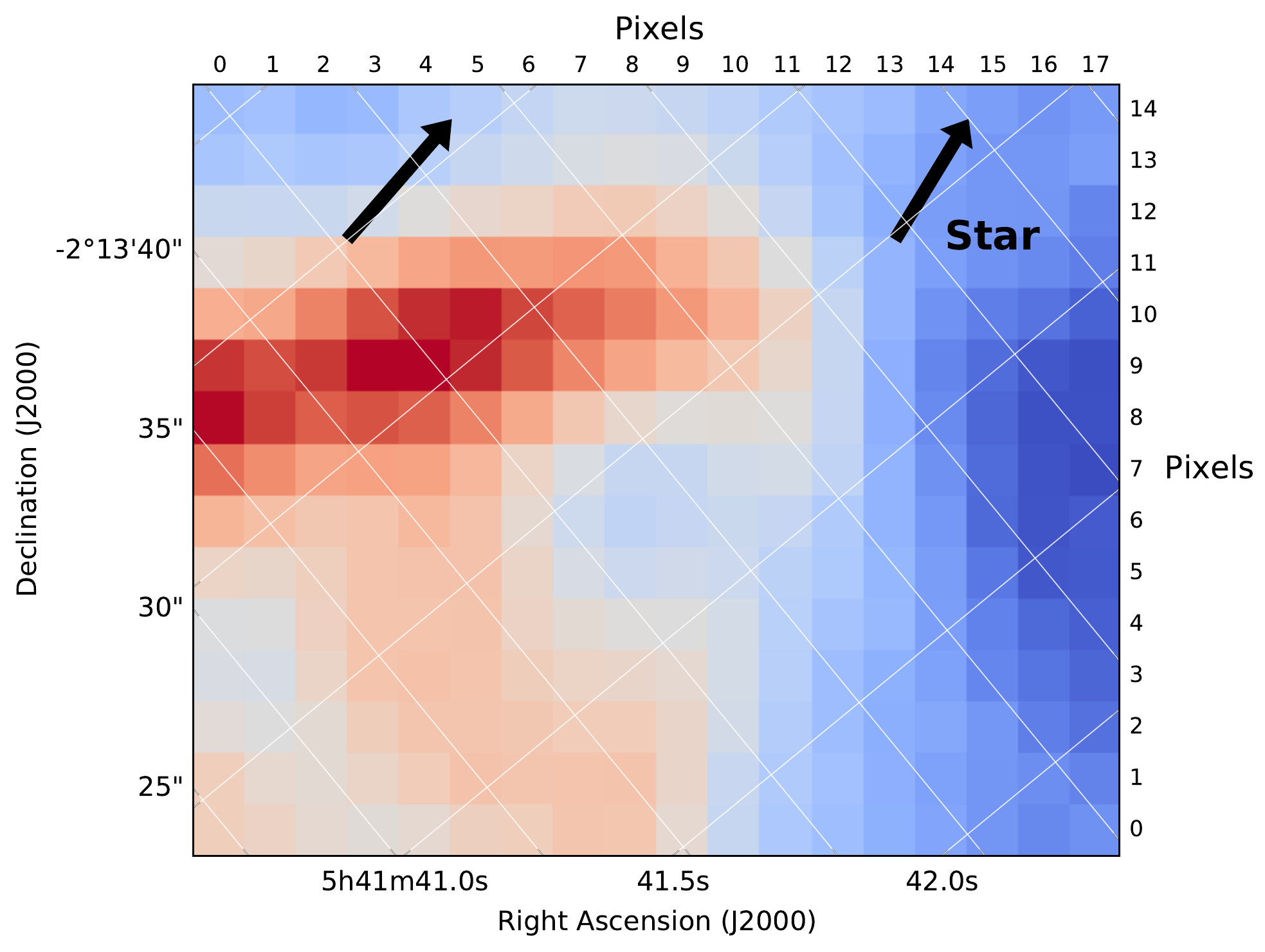}
	\label{fig:astro_2023N}
    }
    \subfigure{
	\includegraphics[height=6cm]{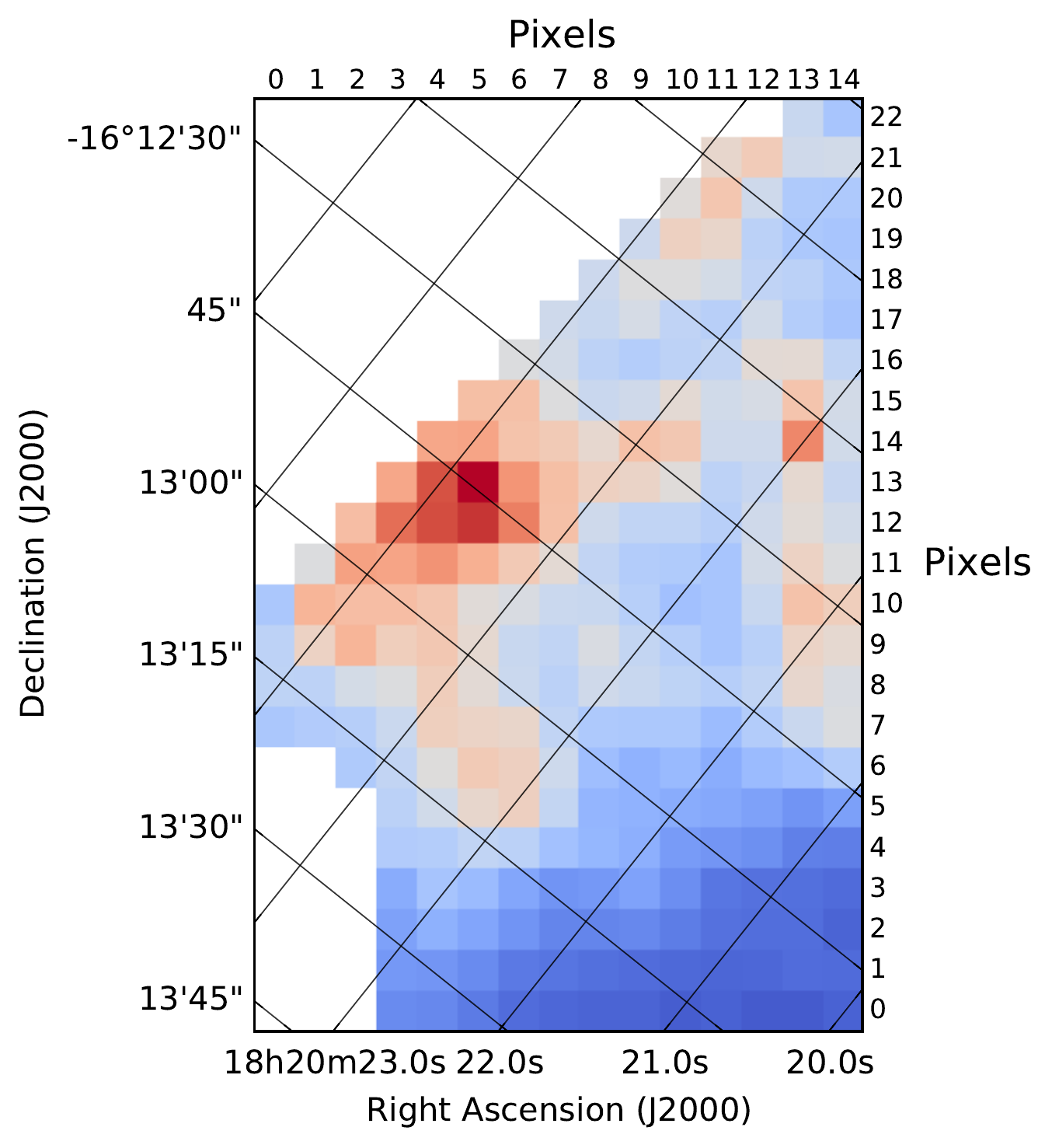}
	\label{fig:astro_M17}
    }
\caption{Astrometry for NGC 2023 North (left) and M17 (right). The normalized flux of the traditional 11.2 \mt band is shown in each map (cf. Figs.~\ref{fig:map_2023N} and ~\ref{fig:map_M17}). The illuminating star for NGC 2023 North is HD 37903, 83\arcsec~beyond the boundary of the image along the indicated radial vectors (black arrows; two are displayed to illustrate the distance). M17 is more complex, and no arrow is displayed. See \cite{sheffer2013} for details.}
\label{fig:astrometry2}
\end{center}
\end{figure*}

Spectroscopic maps were obtained using the Infrared Spectrograph (IRS; \citealt{houck2004}) on-board the \textit{Spitzer} Space Telescope \citep{werner2004}. These data, spanning 10-20 \mt at resolving power R$\sim$600, were obtained from the NASA/IPAC \textit{Spitzer} Heritage Archive\footnote{\url{http://sha.ipac.caltech.edu/applications/Spitzer/SHA/}}. The spectral maps included in this work are summarized in Table~\ref{table:target_properties}. Astrometry for NGC 7023 and NGC 2023 South is presented in Fig.~\ref{fig:astrometry}, and for NGC 2023 North and M17 in Fig.~\ref{fig:astrometry2}.

\begin{table*}
\caption{\label{tables:objects}Target properties \& observation log}
\begin{center}
\begin{tabular}{l c c c c c c c c}
\hline\hline
Object            & Distance & Exciting Star & Field of view$^a$          & Field of view$^a$      	 & AORkey$^b$ & References \\
                  &  [kpc]   & spectral type & \arcmin $\times$ \arcmin & pc $\times$ pc  	 	 &  & \\
NGC 7023          & 0.43     & B2.5          & 1.13 $\times$ 0.94           & 0.14 $\times$ 0.12 	 & 3871232  & 1,2\\
NGC 2023 South    & 0.35     & B1.5          & 1.24 $\times$ 0.86           & 0.13 $\times$ 0.09	 & 14033920 & 3\\
NGC 2023 North    & 0.35     & B1.5          & 0.72 $\times$ 0.56           & 0.07 $\times$ 0.06 	 & 26337024 & 3\\
M17-SW				  & 1.98     & O4$^c$              & 3.27 $\times$ 1.67			   & 1.90 $\times$ 0.96 	 & 11543296 & 4,5,6,7\\
\hline
\end{tabular}
\end{center}

\centering
$^a$ Field of view of the SH maps. $^b$ The AORkey uniquely identifies \textit{Spitzer} Space Telescope observations. $^c$ The major exciting star is thought to be CEN 1, a double O4-type star \citep{chini1980,hoffmeister2008}.
\tablerefs{(1) \cite{sellgren2007}; (2) \cite{rosenberg2011}; (3) \cite{peeters2012}; (4) \cite{xu2011}; (5) \cite{hoffmeister2008}; (6) \cite{chini1980}; (7) \cite{sheffer2013}
}

\bigskip
\label{table:target_properties}
\end{table*}

\subsection{Data reduction}

The NGC 2023 North and South SH maps were previously presented by \cite{peeters2012} and \cite{shannon2015}. These data were processed by the \textit{Spitzer} Science Center (pipeline version S18.7). Further processing was accomplished with the \texttt{CUBISM} tool \citep{smith2007cubism}, including coaddition and bad pixel cleaning. For the purpose of spectral extraction, a $2\times 2$-pixel aperture was stepped across each map. This ensured that the extraction apertures matched the point-spread function, removing non-independent pixels. Further details of the reduction process can be found in \cite{peeters2012}. A similar approach was applied in the reduction of the SH maps of NGC 7023 and M17. The map of NGC 7023 has been previously analyzed by \cite{rosenberg2011,berne2012,boersma2013,shannon2015} and \cite{sellgren2007}. \textit{Spitzer}/IRS observations of M17-SW have been previously examined by \cite{povich2007} and \cite{sheffer2013}.

Extinction is small in NGC 2023, as are the variations in extinction, as concluded by \cite{peeters2016} and \cite{stock2016}, with A$_k$ values on the order of 0.1. We thus do not correct for it in this source. The extinction in NGC 7023 and M17 was investigated by \cite{stock2016}. These authors computed the extinction with the iterative Spoon method \citep{spoon2007,stock2013} using \textit{Spitzer}/IRS-SL data. Regarding NGC 7023, \cite{stock2016} found significant extinction (A$_k\sim2$) in the lower left corner of the map, as also reported by \cite{boersma2013}. The small amount of extinction in the rest of the map, combined with the fact that the gradients of the extinction curve in the 11.0-11.6 and 12.3-13 \mt regions are small, leads to a change in profile shapes of the 11.2 and 12.7 \mt features at approximately the 4\% level towards NGC 7023. In contrast, the extinction towards M17 reaches a maximum A$_k$ value of 1.48 in our field, with typical values near unity. Therefore, we dereddened our spectra using the Chiar \& Tielens (2006) extinction law. Further discussion and analysis on this topic is presented in Section~\ref{subsec:127_over_112}, including its effect on the 12.7/11.2 band strength ratio.

\subsection{Continuum estimation}

\begin{figure}
	\centering
	\includegraphics[width=1.0\linewidth]{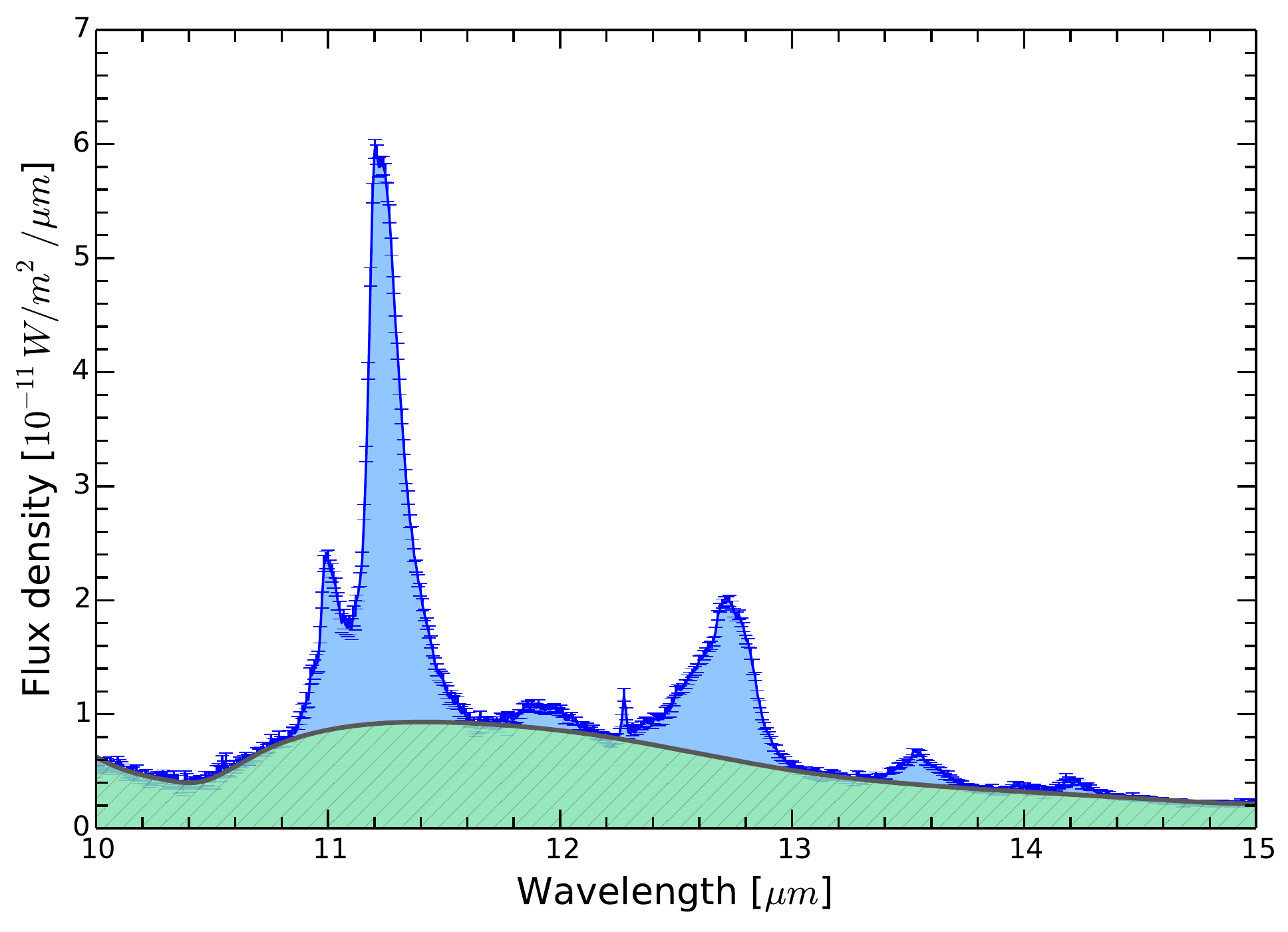}
	\caption{
	Sample spectrum from NGC 7023. A spline continuum is defined by the green hatched area. The prominent 11 and 12.7 \mt emission complexes are visible, in addition to a variety of smaller features at (e.g.,) 12.0, 13.5, 14.2 \mt and molecular hydrogen emission at 12.3 \m.
    }
	\label{fig:spectrum}
\end{figure}

A local spline continuum was chosen to isolate the PAH emission features. A series of anchor points were chosen between 10-20 \mt (Fig.~\ref{fig:spectrum}) to define the spline. This type of continuum determination has been performed several times in the literature and is chosen here for the purposes of comparison (e.g., \citealt{vankerckhoven2000,hony2001,peeters2002,galliano2008b}). The PAH band profiles and fluxes derived from the spline method are not very sensitive to the precise position of the anchor points, depending on the apparent smoothness of the underlying continuum (i.e. spectra with, e.g., undulating continua will lead to less precise PAH band flux measurements). We estimate the influence of the continuum choice on our 11.2 and 12.7 \mt band fluxes to be at the 5\% level.

\section{Spectral profiles}
\label{sec:specprof}

We focus here specifically on the PAH emission bands at 11.0, 11.2 and 12.7 \m. The former two bands we will generally refer to as the ``11 \m" emission complex, and the latter as the 12.7 \mt emission complex. The 12.0 \mt PAH band is excluded as it does not exhibit spectral blending with the 12.7 \mt emission.

\begin{figure*}
	\centering
        \includegraphics[width=\linewidth] {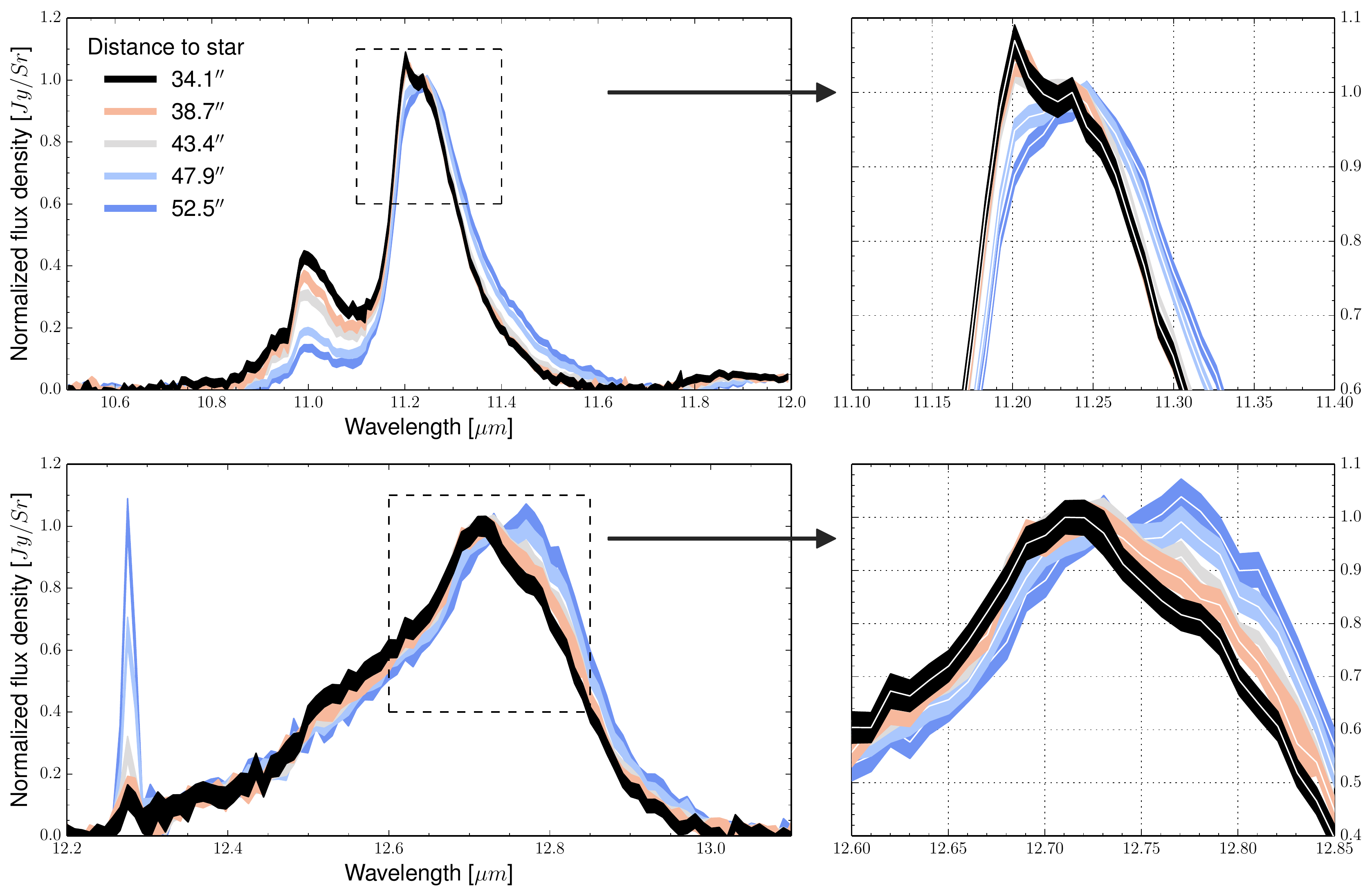}
\caption{
The continuum-subtracted emission in NGC 7023 of the 11 \mt emission complex (top) and the 12.7 \mt emission complex (bottom). The colour-coding represents different stellar distances. On the right is an expanded view of the emission peak, identified by the dashed rectangle. The width of each line represents the $\pm1\sigma$ uncertainty on the flux density measurements.
}
\label{fig:raw}
\end{figure*}

The variability of the 11 \mt and 12.7 \mt emission complexes in NGC 7023 is examined in Fig.~\ref{fig:raw}. A representative set of pixels are chosen from the map such that each position is a different distance from the illuminating source.

The relative strengths of the 11.0 and 11.2 \mt emission bands show significant variations across the map (upper panel). As stellar distance decreases, four effects are simultaneously visible: the 11.0 \mt band strength relative to the 11.2 \mt emission increases monotonically; the peak position of the 11.2 \mt band moves to shorter wavelengths (from class (A) to class A(B); see also \citealt{boersma2012,boersma2013}); a small, narrow feature at 11.20 \mt appears; and the red wing of the 11 \mt complex either shifts towards shorter wavelengths or decreases in intensity (also reported by \cite{boersma2012,boersma2013} for Orion and NGC 7023, respectively).

For these same positions, we inspect the behaviour of the 12.7 \mt complex (Fig.~\ref{fig:raw}, lower panel). As the stellar distance decreases, there is a clear transition of the 12.7 \mt peak position towards shorter wavelengths---from approximately 12.77 to 12.71 \m. The red wing also decreases in intensity (or shifts to shorter wavelengths) during this transition. In addition, there is a difference in the blue wing of the 12.7 \mt complex between the map positions: in the range 12.5 - 12.7 \m, positions closer to the star display greater emission intensities than those further away. It does not appear that the entire emission complex is shifting, as the offsets of the red wing, peak position and blue wing are not consistent. Likewise, the emission blueward of 12.5 \mt are identical in each position.

The other sources exhibit only minor variations. M17 displays a similar peak transition in the 11.2 \mt band, while NGC 2023 North and South show little peak variation in the 11 \mt complex. At 12.7 \m, NGC 2023 South and M17 have small variations in the peak profile and red wing, but not as clearly as for NGC 7023. The North map of NGC 2023 has no discernible 12.7 \mt variations.

The observed sub-structure of the emission complexes in NGC 7023 suggests that subtle and possibly important clues about PAH sub-populations may be accessible.

\section{Methods for analyzing the PAH bands}
\label{sec:methods}

We now discuss traditional methods and a newly proposed method for analyzing the 11 and 12.7 \mt PAH bands.

\subsection{The traditional approach}
\label{sec:tradition}

The 11 \mt emission complex consists of two components: a strong, broad feature near 11.2 \mt and a weaker band near 11.0 \m. They are generally blended together near 11.1 \m. There are multiple ways in which to measure the fluxes in these two emission bands. The simplest method is to integrate the 11 \mt complex shortward of 11.1 \mt (representing the 11.0 \mt band) and longward of 11.1 \mt (representing the 11.2 \mt band). A more complex method involves using Gaussian components to help disentangle the spectral blend (see, e.g., \citealt{stock2014}, Peeters et al. 2015, ApJ, submitted). In this case, one simultaneously fits a Gaussian to the 11.0 \mt band and a Gaussian to the blue wing of the 11.2 \mt complex. The 11.0 \mt flux is then the flux contained within the fitted Gaussian at 11.0 \m. The 11.2 \mt band flux is measured by subtracting the 11.0 \mt Gaussian from the original spectrum and integrating the remainder. A good single Gaussian fit to the 11.2 \mt band is not possible due to the asymmetric red wing (\citealt{pech2002,vandiedenhoven2004}). A third method to separate the 11.0 and 11.2 \mt emission was introduced by \cite{boersma2012}. The authors used a five-component Gaussian decomposition in which one component was responsible for the 11.0 \mt emission. The sum of the remaining components then represented the 11.2 \mt emission. A fourth possibility is to fit the 11 \mt emission with two Drude profiles, an approach used by the PAHFIT tool \citep{smith2007}.

For the 12.7 \mt band, one measurement method (besides direct integration) was presented by \cite{stock2014}. Their approach uses the average 12.7 \mt emission profile from \cite{hony2001} as a template. This template is scaled to the data in the range 12.4-12.7 \m, where only PAH emission is expected. There is frequently an adjacent [Ne~\textsc{II}] 12.81 \mt atomic line and H$_2$ emission at 12.3 \m, which prevents scaling beyond this spectral window. After fitting the PAH template and subtracting its profile, only atomic and/or molecular emission lines remain. These are fit with Gaussian functions in accordance with the instrumental spectral resolution of the data. Afterwards, the atomic and molecular lines are subtracted from the original 12.7 \mt emission complex, leaving only the 12.7 \mt PAH emission. As with the 11 \mt emission, another possible 12.7 \mt decomposition is to fit the band with two Drude profiles (the PAHFIT approach).

To account for the diversity of the profile variations (see Section~\ref{sec:specprof}), we have introduced a new decomposition method which we will explore in detail.

\subsection{A new decomposition}
\label{sec:decomp11}

We attempted to fit the 11 and 12.7 \mt complexes with a variety of Gaussians and Lorentzians. We used the non-linear least squares fitting tool MPFIT \citep{markwardt2009}, which is an implementation of the Levenberg-Marquardt algorithm \citep{more1978}. We compute the reduced-$\chi^2$ of the fit for each pixel in our spectral cubes. A histogram of all pixels in each cube is prepared, which is used to evaluate the overall fit. The best reduced-$\chi^2$ values resulted from fitting the 11 \mt complex with five Gaussians and the 12.7 \mt complex with four. To determine their nominal parameters we applied an iteration method. Initially, the components were permitted to move, as long as they did not overlap. Their widths were also allowed to vary. After allowing several ``free runs" we observed that some of the parameters were converging on particular values (those areas of the spectrum with less complex structure). We fixed these converging parameters and recomputed our fits. The process was iterated in this manner until all parameters had converged. The final fits are thus obtained with fixed peak positions and full-width at half-maximum (FWHM) values while only the peak intensity of each component was allowed to vary.

It is important to emphasize that \textit{the decompositions we adopt are arbitrary}. We do not suggest that they reflect a ``true decomposition" or any a priori knowledge. Rather, this method is applied to understand what (if anything) may be learned about the PAH bands through simple fitting.

\paragraph{The 11 \mt complex} The best fits resulted from a five-component Gaussian fit. The Gaussians were fixed to the following positions: 11.021 \mt (FWHM: 0.066 \m), 11.000 \mt (0.021 \m), 11.199 \mt (0.021 \m), 11.320 \mt (0.118 \m) and 11.245 \mt (0.055 \m). An example of the decomposition we adopted is shown in Fig.~\ref{fig:112_3pos}. We see that the 11.0 \mt emission is determined by two components, and the 11.2 \mt emission by three components. The two components of the 11.0 \mt emission can be qualitatively interpreted as an underlying broad component (centered near 11.0 \m) and a narrower symmetric feature placed upon it. The 11.2 \mt profile displays a strong Gaussian feature centered near 11.25 \mt with an accompanying red wing (out to 11.6 \m). The slight asymmetry of the 11.2 \mt peak, namely on the short-wavelength edge (near 11.20 \m), is coincident with the fifth component.

\paragraph{The 12.7 \mt complex} We applied the same methodology to the 12.7 \mt emission complex and found that the best statistics resulted from a four-component decomposition. A sample decomposition is presented in Fig.~\ref{fig:127_3pos}. One component of the fit is linked to the broad blue wing, upon which a weaker, narrower component is placed. The other two components compete to fit the peak position of the band and the spectral profile of its long-wavelength wing. The adopted central wavelengths of our components are as follows: 12.55 \mt (FWHM: 0.160 \m), 12.54 \mt (0.035 \m), 12.72 \mt (0.090 \m) and 12.78 \mt (0.080 \m).

\section{Results}
\label{sec:results}

\subsection{Spectral characteristics of the fit}

We present our decompositions of the 11 and 12.7 \mt emission complex in Figs.~\ref{fig:112_3pos} and ~\ref{fig:127_3pos}, respectively, for a choice of three locations in NGC 7023. These positions are chosen along a radial vector emanating outward from the central star. The positions are 34\arcsec, 43\arcsec~and 53\arcsec~distant from the star, respectively.

\subsubsection{The 11 \mt emission}

\begin{figure}
\begin{center}
    \subfigure{
        \includegraphics[width=\linewidth] {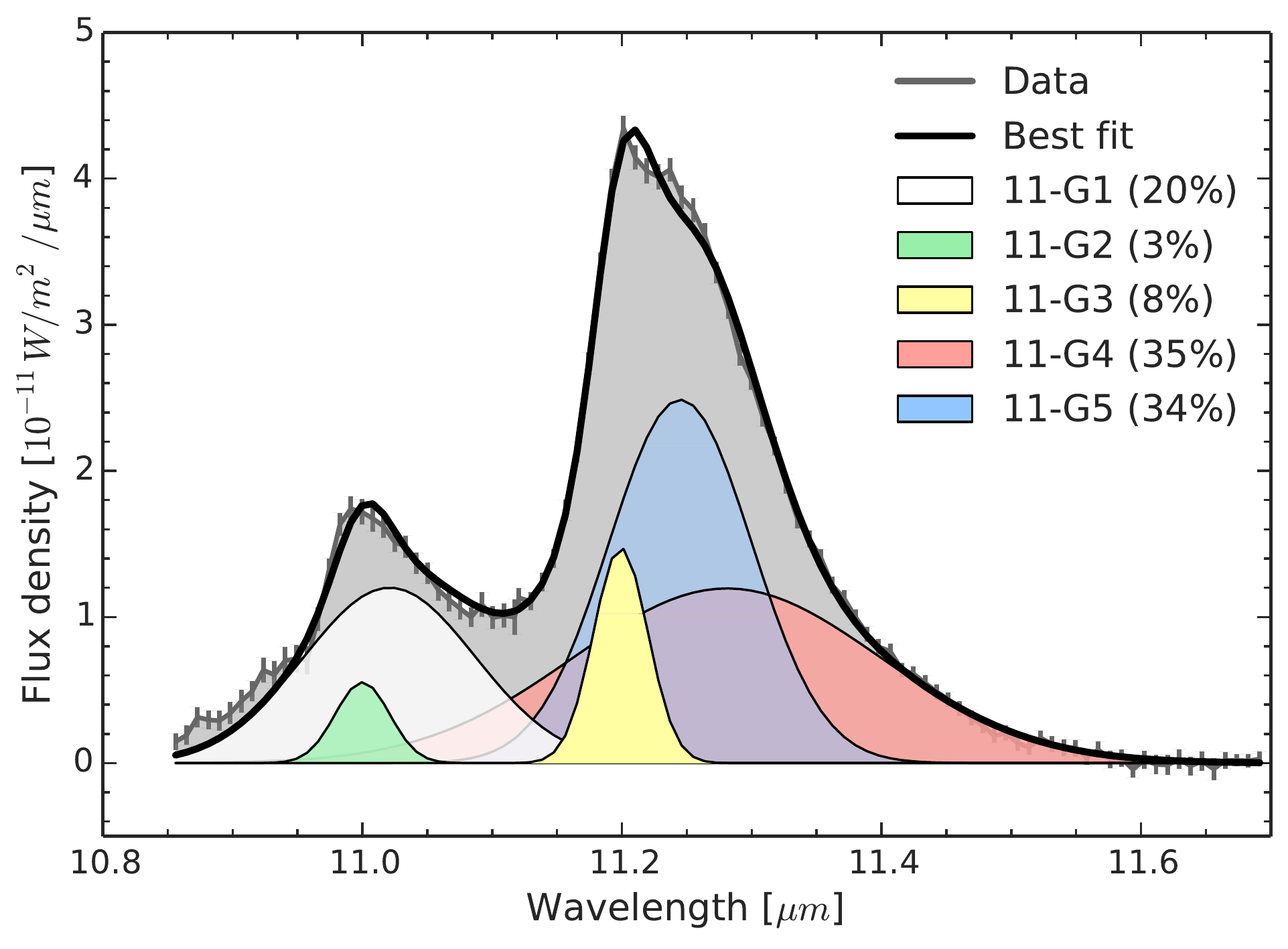}
    } \\
    \subfigure{
        \includegraphics[width=\linewidth] {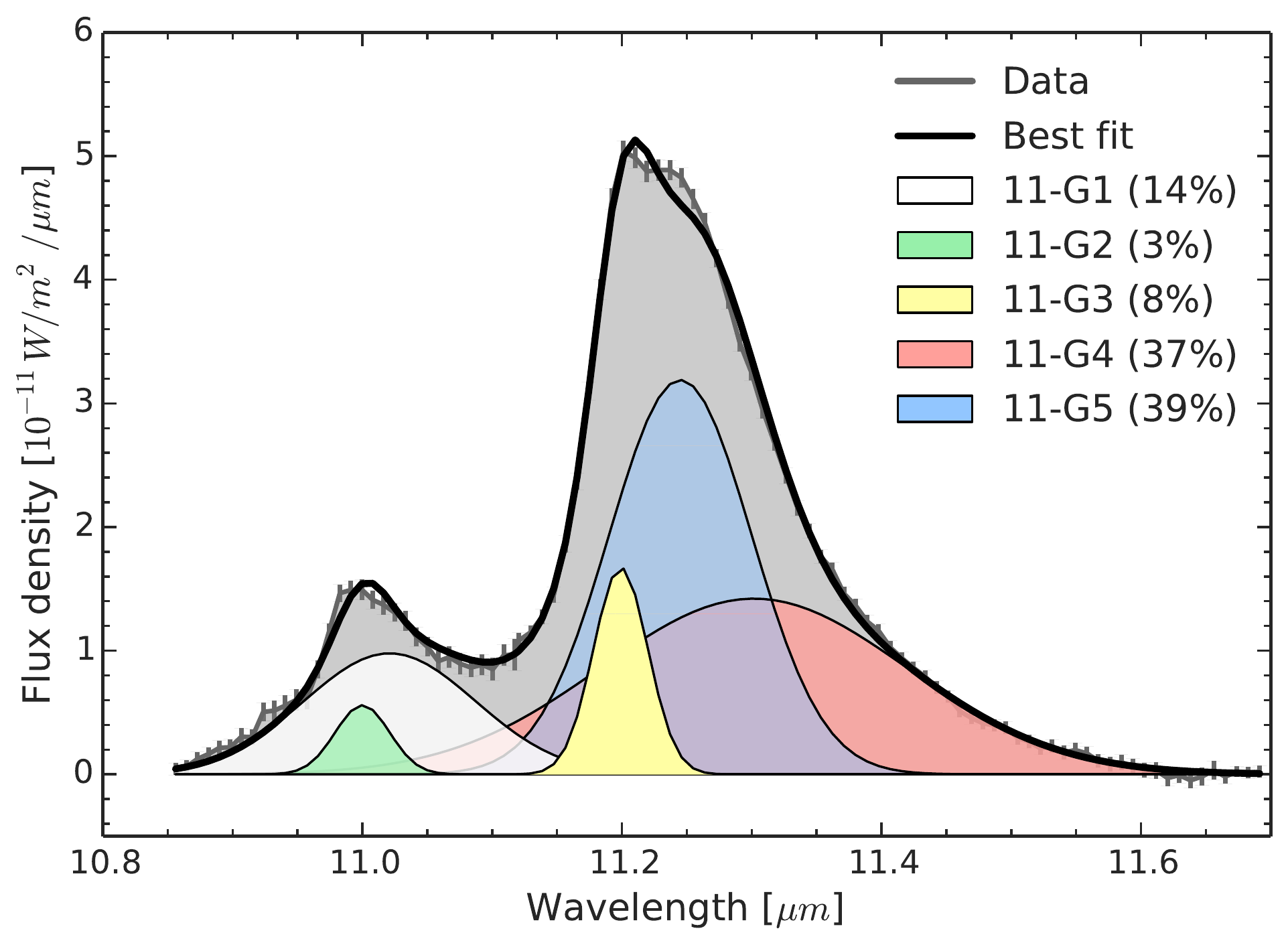}
    } \\
    \subfigure{
        \includegraphics[width=\linewidth] {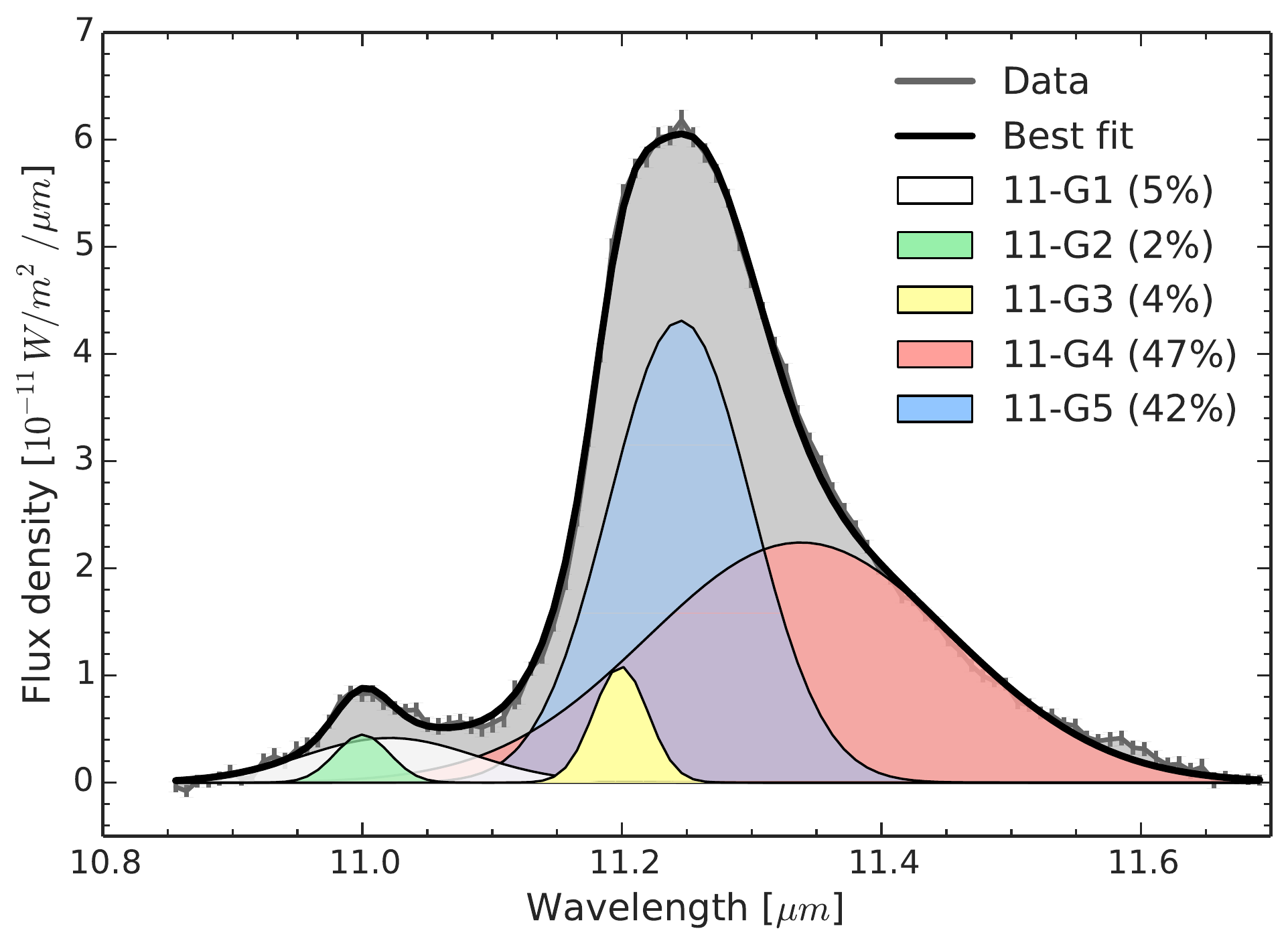}
    }
\caption{The 11 \mt decomposition in NGC 7023 for three positions at varying distances from the central star (34\arcsec~in the upper panel, 43\arcsec~in the middle panel, and 53\arcsec~in the lower panel). These positions correspond to the closest, furthest, and intermediate positions in Fig.~\ref{fig:raw}. The percentages in parentheses indicate the fraction of total flux carried by each component.}
\label{fig:112_3pos}
\end{center}
\end{figure}

Considering the 11 \mt emission first, we see that at the closest location the 11-G1 component dominates the 11.0 \mt emission (Fig.~\ref{fig:112_3pos}, upper panel). Here, the 11-G2 component contributes weakly. At further distances, however, the flux of the 11-G1 component decreases significantly, while the flux of the 11-G2 component is relatively unchanged. At the furthest distance, the peak flux density of the 11-G2 component slightly exceeds that of the 11-G1 component. The 11.2 \mt emission profile varies across the three map positions. Near the star, the 11.2 \mt peak is asymmetric and centered near 11.20 \m. Far from the star, we see that the peak of the 11.2 \mt emission has shifted to $\sim$11.25 \mt and it is now more symmetric. Note that the fit has a slight mismatch on the peak emission at the closer positions.

The 11-G3 component decreases in intensity with distance, similar to that of the 11-G1 emission. The 11-G3 component appears to be at least partially responsible for the asymmetry of the peak 11.2 \mt emission at positions near the star. The other two components, 11-G4 (which traces the red-shaded wing) and 11-G5 (which contains the bulk of the flux at the traditional position of 11.2 \m) increase in flux density with increasing distance.

\subsubsection{The 12.7 \mt emission}

\begin{figure}
\begin{center}
    \subfigure{
        \includegraphics[width=\linewidth] {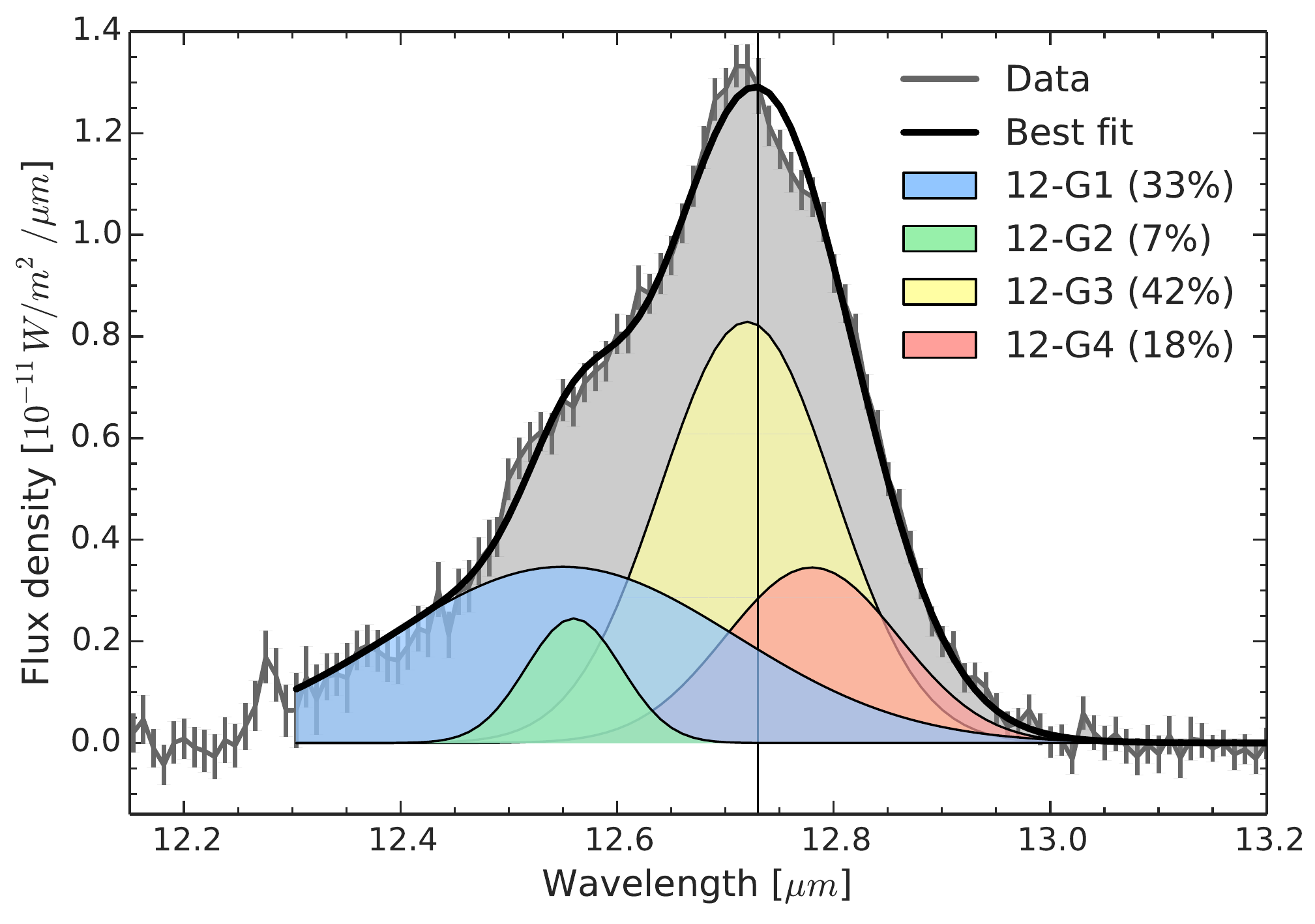}
    } \\
    \subfigure{
        \includegraphics[width=\linewidth] {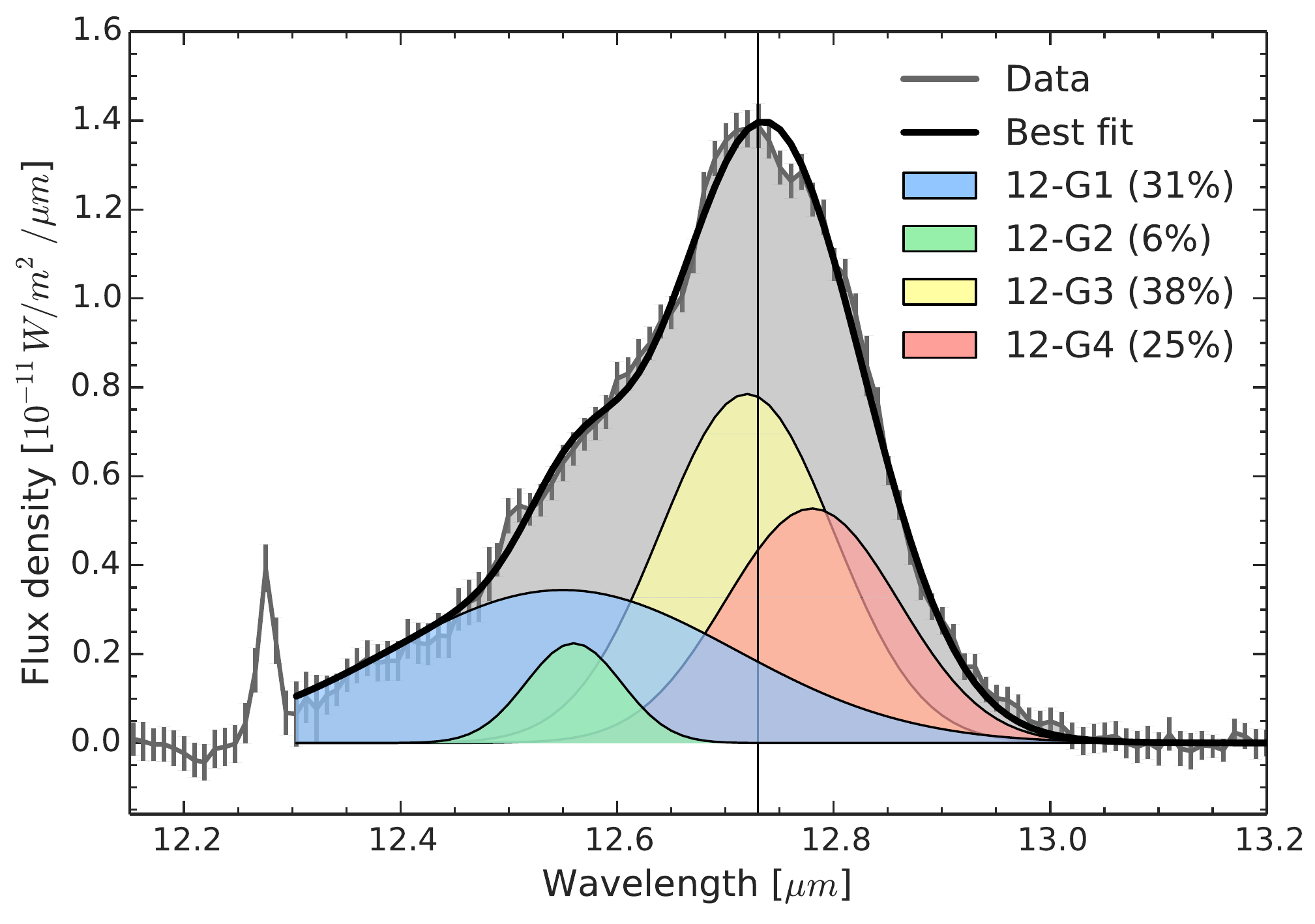}
    } \\
    \subfigure{
        \includegraphics[width=\linewidth] {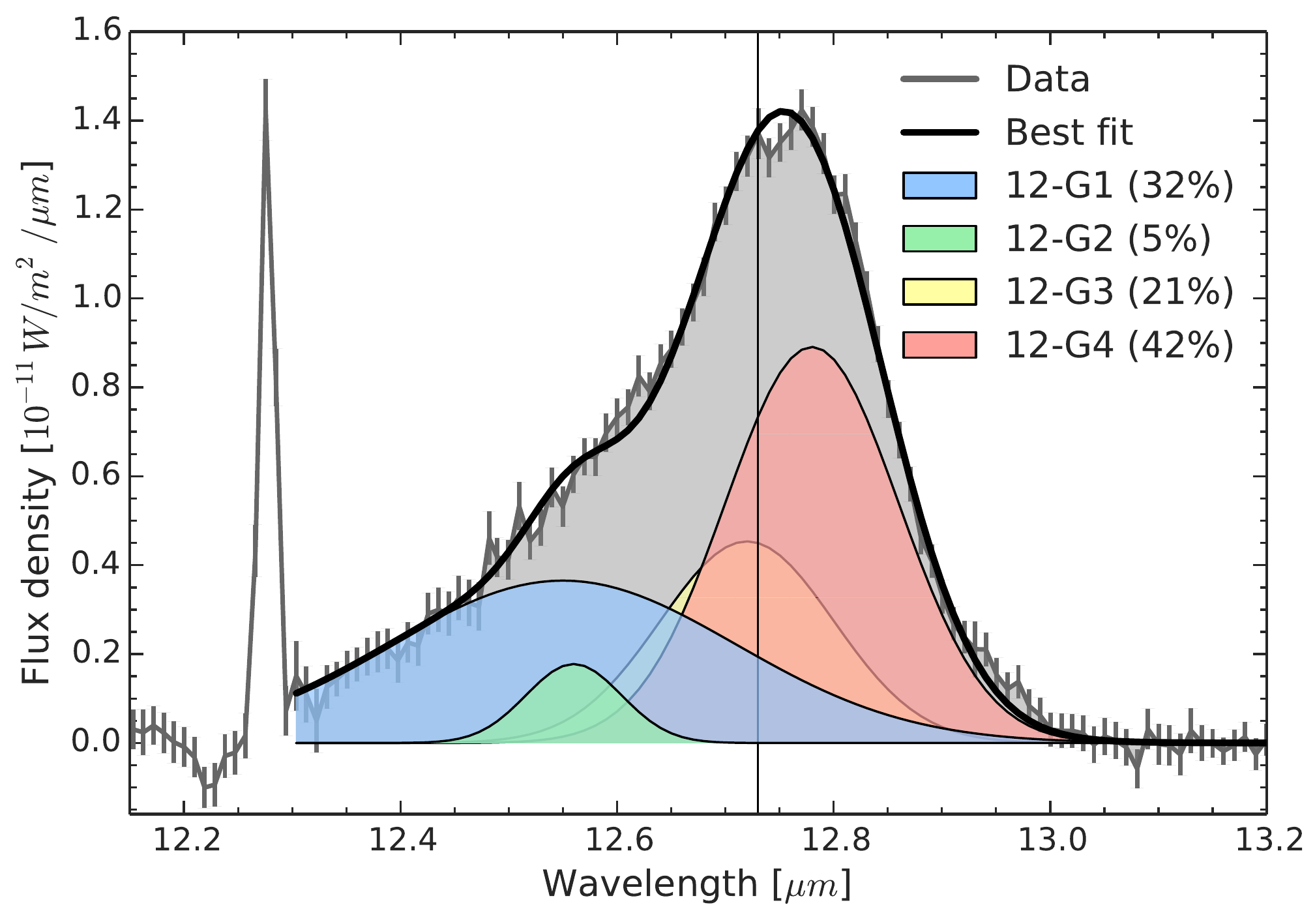}
    }
\caption{The 12.7 \mt decomposition in NGC 7023 for three positions at varying distances from the central star (34\arcsec~in the upper panel, 43\arcsec~in the middle panel, and 53\arcsec~in the lower panel). These positions correspond to the closest, furthest, and intermediate positions in Fig.~\ref{fig:raw}. The percentages in parentheses indicate the fraction of total flux carried by each component. The vertical line at 12.73 \mt is to guide the eye.}
\label{fig:127_3pos}
\end{center}
\end{figure}

In Fig.~\ref{fig:127_3pos} we see that at the position nearest the star the peak of the 12.7 \mt complex lies blueward of 12.75 \mt (which we use as a reference). At this location, the flux of 12-G3 is clearly greater than that of 12-G4.  We observe strong asymmetry of the 12.7 \mt peak at this position. Note the fit cannot completely reproduce the peak shape observed. At the position furthest from the central star (lower panel), however, the peak is now located redward of 12.75 \m. The flux of the 12-G4 component now exceeds that of the 12-G3 component considerably, indicating that it is the relative strengths of the 12-G3 and 12-G4 components that determines the peak position of the 12.7 \mt emission. At an intermediate distance from the central star in NGC 7023 (middle panel), we see that 12-G3 and 12-G4 are of roughly equal strength, and that the 12.7 \mt emission has a peak that is an intermediate between the extremes of the upper and lower panels. The broad component in this decomposition, 12-G1, is generally unchanged across the three map positions. The 12-G2 component  however decreases in relative intensity as one moves to further distances from the central star.

\subsection{Spatial morphology}
\label{sec:results_maps}

\subsubsection{The 11 \mt decomposition}
\label{sec:11decomp}

Starting with NGC 7023, we present spectral maps of the five components of the 11 \mt decomposition in Fig.~\ref{fig:map_7023}. We have included maps of the traditional 11.0 and 11.2 \mt decomposition, defined here as direct integration, for comparison (see Section~\ref{sec:tradition}). We observe that the distribution of the 11-G1 component is extremely similar to that of the 11.0 \mt band. Recall that the 11-G1 component is the broad underlying feature centered near 11.0 \m. The 11-G2 component, which is the smaller feature perched on top of the 11-G1 plateau, shows a spatial distribution that is intermediate between that of the 11.0 and 11.2 \mt emission. This is also true for the 11-G3 component, which appears to be intermediate in morphology between the 11.0 and 11.2 \mt bands. The 11-G4 and 11-G5 components are both coincident with the 11.2 \mt emission.

Similar results are found in NGC 2023 South (Fig.~\ref{fig:map_2023S}). The map of the 11-G1 component's flux is an extremely good match to that of the traditional 11.0 \mt component. The 11-G2 and 11-G3 components are also intermediate in morphology between that of the 11.0 and 11.2 \mt bands. However, it is clear that in NGC 2023 South the 11-G3 component is closer in spatial distribution to that of the 11.0 \mt emission and the 11-G2 component is closer in spatial distribution to that of the 11.2 \mt emission. The 11-G4 and 11-G5 components again match the 11.2 \mt emission, though the 11-G4 component is less extended in comparison.

The spectral maps of NGC 2023 North are presented in Fig.~\ref{fig:map_2023N}. We observe that the 11.0 \mt band and the 11-G1 components are similar, each peaking along a vertical slice in the map. Additionally, the 11-G3, 11-G4 and 11-G5 components exhibit similar morphologies, each peaking along a common horizontal strip. The 11-G2 component appears to share the peak positions of these two groups, forming an almost ``L" shape from the horizontal and vertical peak PAH zones.

In M17 (Fig.~\ref{fig:map_M17}), there is generally very little variation between the traditional 11.0 and 11.2 \mt emission maps. They appear to peak in the same position in the map. However, there is one apparent difference: there is 11.2 \mt emission in the upper corner, forming a small ridge along the boundary of the field of view. We will hereafter refer to this feature as the M17 ``spur.'' Using this as a distinguishing characteristic, we observe that the 11.2 \mt band, the 11-G4 component and the 11-G5 component all have emission in this region. Conversely, the 11.0 \mt band and the 11-G1, 11-G2 and 11-G3 components all have very little emission in comparison. We also observe that the 11-G2 component is more extended than the 11-G1 and 11-G3 emission, and is closer in morphology to that of the 11-G4 and 11-G5 bands. This suggests that the 11-G1 and 11-G3 bands are more closely related to the 11.0 \mt emission than the 11-G2 component.

In brief summary, the emission of the 11-G1 component closely spatially matches the 11.0 \mt band. The maps of the 11-G2 and 11-G3 components are a mixture of the 11.0 and 11.2 \mt maps, while the 11-G4 and 11-G5 distributions are well-matched to that of the traditional 11.2 \mt band.

\subsubsection{The 12 \mt decomposition}
\label{sec:12decomp}

\begin{figure*}
	\centering
\includegraphics[width=0.9\linewidth, trim={0cm 9.5cm 0cm 0cm}, clip]{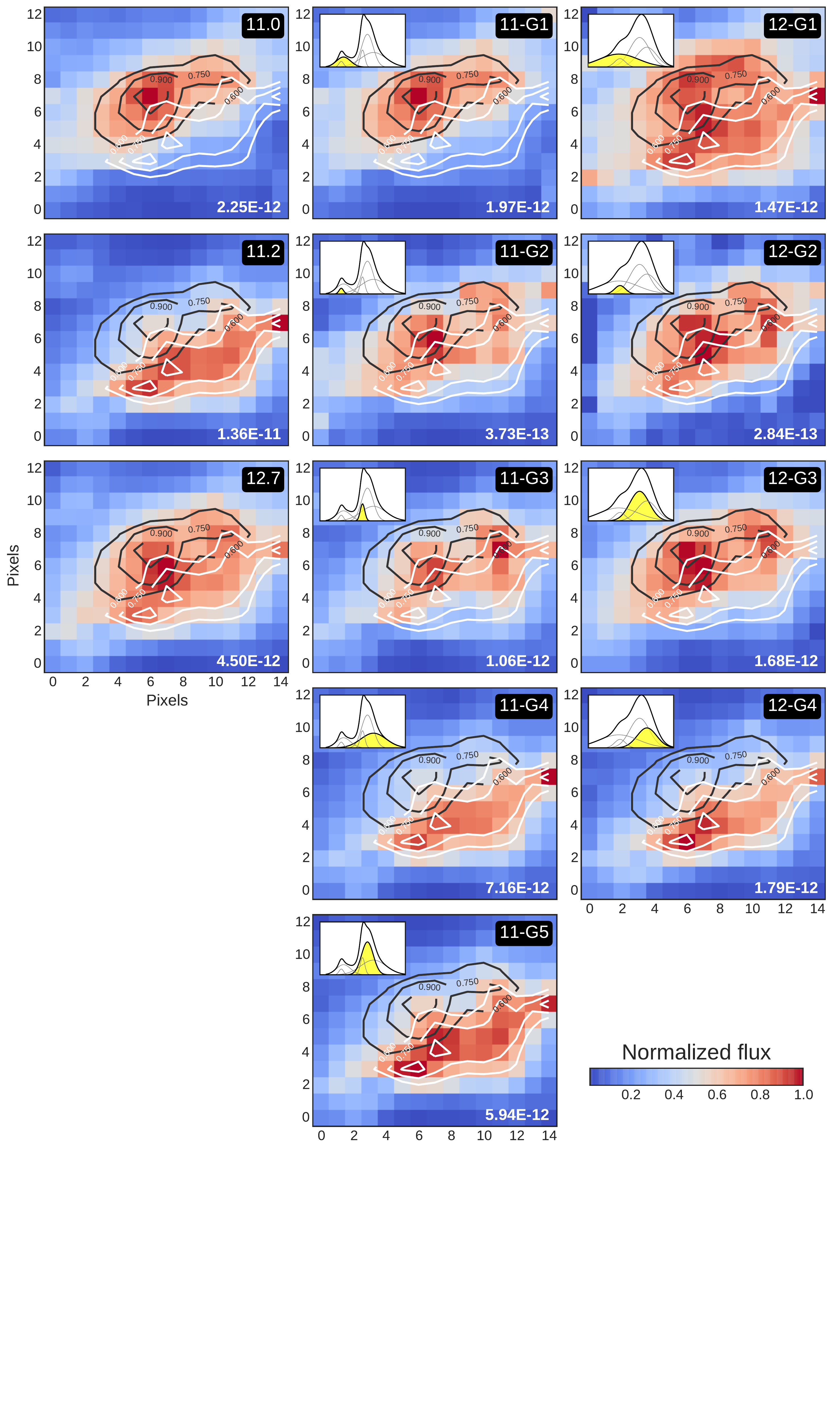}
	\caption{PAH emission in NGC 7023. The left column displays the 11.0, 11.2 and 12.7 \mt emission measured using the traditional methods. The middle column displays the flux of each component in the 11 \mt decomposition, and the right column displays the emission for the 12.7 \mt decomposition. Each map is normalized, with the normalization constant given in the lower-right corner. The name of each band is identified by the black rectangle and the inset figures identify which component is being plotted (highlighted in yellow). The black and white contours trace the traditional 11.0 and 11.2 \mt emission, respectively.
	}
	\label{fig:map_7023}
\end{figure*}

In Fig.~\ref{fig:map_7023} we present maps of the four components of the 12.7 \mt decomposition. We again include maps of the 11.0 and 11.2 \mt emission bands as measured with the traditional decomposition (see Section~\ref{sec:tradition}). As shown in Fig.~\ref{fig:127_3pos}, it is the 12-G3 and 12-G4 components whose relative strengths determine the position of the 12.7 \mt peak. The 12-G3 component of the 12.7 \mt decomposition has a very similar spatial distribution to that of the 11.0 \mt emission, peaking in generally the same location. The morphologies of the 12-G4 component and the 11.2 \mt emission are likewise very similar. There is generally very little overlap between the 12-G3 and 12-G4 components in these maps. The 12-G2 component generally peaks where the 11.0 \mt emission does, and therefore also the 12-G3 component, but it is clearly more extended than either of these features. The 12-G1 component is the most extended, with emission at both the 11.0 and 11.2 \mt peaks. This is consistent with the findings in Fig.~\ref{fig:127_3pos}, in which the 12-G1 component showed little variation in flux density for three chosen positions in the NGC 7023 map and the 12-G2 component increased in strength when approaching the star.

\begin{figure*}
	\centering
	\includegraphics[width=0.9\linewidth, trim={0cm 13cm 0cm 0cm}, clip]{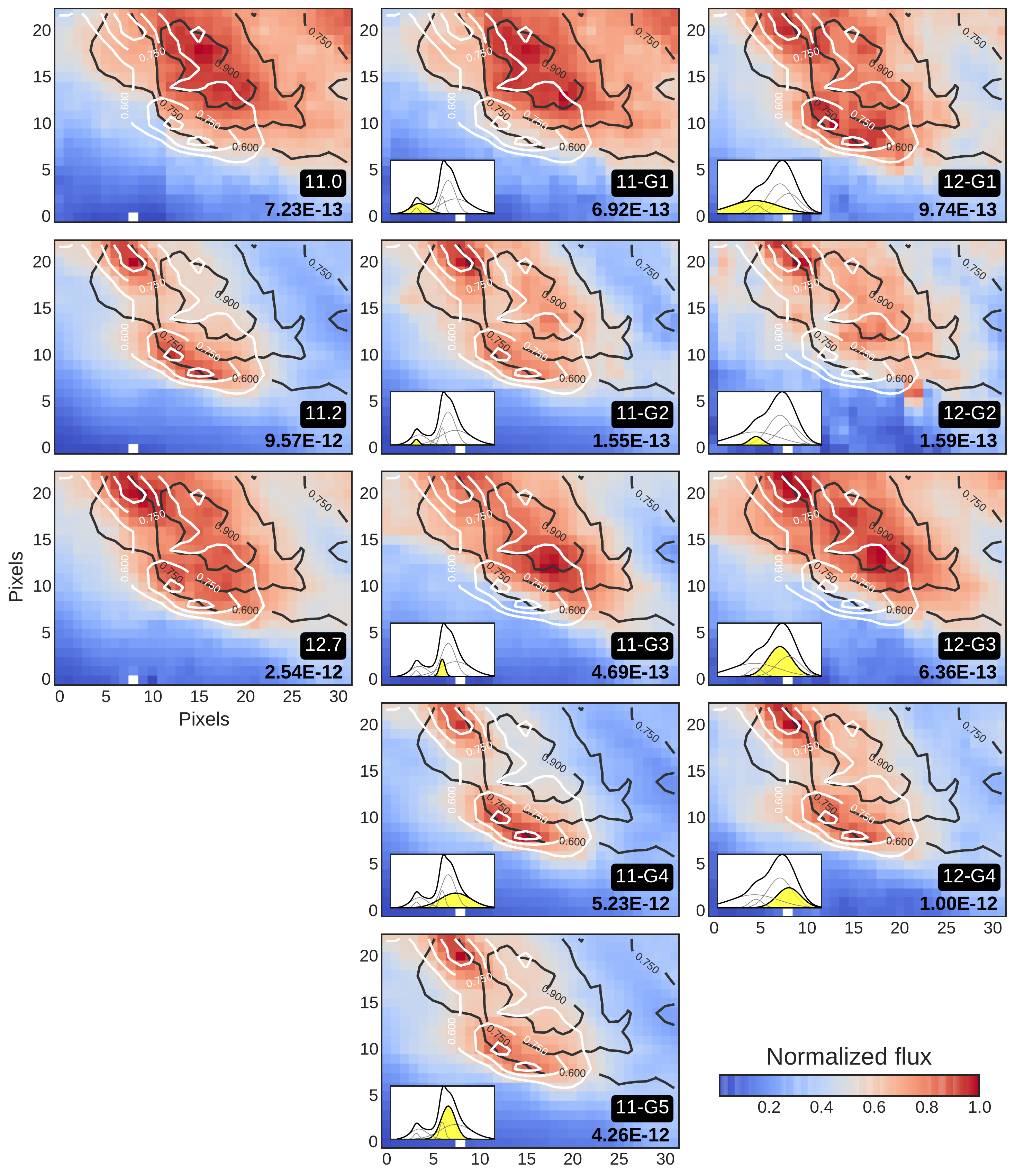}
	\caption{PAH emission in NGC 2023 South. The conventions are the same as those used in Fig.~\ref{fig:map_7023}.
    }
	\label{fig:map_2023S}
\end{figure*}

We perform a similar analysis on the map of NGC 2023 South (Fig.~\ref{fig:map_2023S}). We also observe strong similarities between the spatial distributions 11.0 \mt and 12-G3 fluxes, and the 11.2 \mt and 12-G4 fluxes. The 12-G1 and 12-G2 components both appear to involve a mixture of the spatial emission of the 11.0 and 11.2 \mt bands, which is generally consistent with what was observed in NGC 7023.

The spatial distributions of the emission bands in NGC 2023 North (Fig.~\ref{fig:map_2023N} in the appendix) are difficult to interpret. Very broadly, the peak of the 11.2 \mt emission is coincident with the peak of the 12-G4 component's emission. The 11.0 \mt emission peaks along a vertical line in the map; the 12-G3 component also has strong emission in this region, but there appears to also be significant emission in the same location as that of the 11.2 \mt band (spanning a horizontal zone). The 12-G1 and 12-G2 components are broad and generally overlap the locations in which the 11.0 and 11.2 \mt emission originates. The 12-G1 component has the most extended emission in this map. Spectral maps of the emission in M17 are presented in the appendix in Fig.~\ref{fig:map_M17}. M17 is also difficult to disentangle, but there is roughly a commonality between the 11.2 \mt emission, 12-G4, and 12-G1 components. Likewise, a grouping of the 11.0 \mt emission, 12-G3 and 12-G2 components is observed, as found in NGC 7023.

Summarizing, the 12-G3 emission spatially matches that of the 11.0 \mt band, while the 12-G4 emission component is a close spatial match to the 11.2 \mt emission. The spatial distributions of the 12-G1 and 12-G2 emission maps are clearly a mixture of the 11.0 and 11.2 \mt emission maps.

\section{Discussion}
\label{sec:discussion}

The profile variability of the 11 and 12.7 \mt PAH complexes can be explained within the framework of our adopted decompositions. The asymmetry of the 11.2 \mt peak is determined by the strength of the 11-G3 component, and the asymmetry of the 12.7 \mt peak is determined by the competition between the 12-G3 and 12-G4 components. The spectral maps reveal that some components are spatially coincident while others display distinct spatial morphologies.

\subsection{Structural similarities}

To quantify the morphological similarities of the spectral maps we introduce the structural similarity algorithm of \cite{wang2004}. This is an image processing method to evaluate the similarities between images based on local luminance, contrast and structure. The method produces a structural similarity index (SSIM) to quantify how alike two images are. The SSIM value ranges from -1 to 1, where values approaching 1 represent very similar images (only identical images have an SSIM index of unity). The SSIM index is computed by first comparing sub-regions, or windows, of the two images. The windows are used to compare each portion of the corresponding images, before producing a single number to encapsulate the similarity between the images as a whole. The SSIM index between two images $x$ and $y$ of common size $N \times M$ is defined as follows:
$$ \text{SSIM}(x,y) = \frac{(2\mu_x \mu_y + C_1)(2\sigma_{xy} + C_2)}{(\mu_x^2 + \mu_y^2 + C_1)(\sigma_x^2 + \sigma_y^2 + C_2)} $$
where $\mu$ and $\sigma$ are the mean and variance of each window, respectively. The constants $C_1$ and $C_2$ are variables for preventing instability when the denominator would otherwise approach zero. They are defined as $C_1=(K_1 R)^2$ and $C_2=(K_2 R)^2$, in which $R$ is the dynamic range of the image, and $K_1$ and $K_2$ are canonically 0.01 and 0.03, respectively \citep{wang2004}. We use the structural\_similarity subpackage of the scikit-image Python package \citep{scikit-image} to compute the SSIM values. The default window size of 7 pixels is used, as are the canonical $K_1$ and $K_2$ values.

We present the SSIM values comparing the maps from our decomposition in Table~\ref{tables:ssim7023}. SSIM indices exceeding 0.90 are presented in bold face. Due to the NaNs in the M17 map and the SSIM requirement for rectangular windows we could not analyze the M17 map in this manner.

We find the same results as were determined from visual inspection: the morphologies of the 11-G1 and 12-G3 emission components are well-matched with that of the 11.0 \mt emission; the 11-G4, 11-G5 and 12-G4 emission components are spatially well-matched with the 11.2 \mt emission; and the 11-G2, 11-G3, 12-G1 and 12-G2 emission components are spatially a mixture of the two. Similar conclusions about the 12.7 \mt components are reached when examining correlation plots of band flux ratios (Fig.~\ref{fig:corr}).

\begin{table*}
\caption{\label{tables:ssim7023}Structural similarity indices}
\begin{center}
\begin{tabular}{l|*{13}{c}}

\multicolumn{12}{c}{\emph{NGC 7023}} \\
11.0	&	  	&		&		&		&		&		&		&		&		&		&		\\
11.2	&	0.30	&	  	&		&		&		&		&		&		&		&		&		\\
11-G1	&	\textcolor{red}{\textbf{0.99}}	&	0.30	&	  	&		&		&		&		&		&		&		&		\\
11-G2	&	0.75	&	0.73	&	0.74	&	  	&		&		&		&		&		&		&		\\
11-G3	&	0.69	&	0.78	&	0.69	&	\textcolor{red}{\textbf{0.93}}	&	  	&		&		&		&		&		&		\\
11-G4	&	0.24	&	\textcolor{red}{\textbf{0.98}}	&	0.24	&	0.66	&	0.71	&	  	&		&		&		&		&		\\
11-G5	&	0.32	&	\textcolor{red}{\textbf{0.99}}	&	0.32	&	0.75	&	0.79	&	\textcolor{red}{\textbf{0.96}}&	  	&		&		&		&		\\
12-G1	&	0.63	&	0.68	&	0.64	&	0.74	&	0.71	&	0.67	&	0.69	&	  	&		&		&		\\
12-G2	&	0.75	&	0.71	&	0.75	&	\textcolor{red}{\textbf{0.93}}	&	\textcolor{red}{\textbf{0.90}}	&	0.64	&	0.74	&	0.74	&	  	&		&		\\
12-G3	&	0.88	&	0.60	&	0.88	&	\textcolor{red}{\textbf{0.94}}	&	\textcolor{red}{\textbf{0.91}}	&	0.52	&	0.63	&	0.75	&	\textcolor{red}{\textbf{0.94}}	&	  	&		\\
12-G4	&	0.22	&	\textcolor{red}{\textbf{0.96}}	&	0.22	&	0.65	&	0.69	&	\textcolor{red}{\textbf{0.96}}	&	\textcolor{red}{\textbf{0.95}}	&	0.60	&	0.64	&	0.51	&	  	\\
\midrule
	&   11.0&   11.2&  11-G1&  11-G2&  11-G3&   11-G4&  11-G5&   12-G1&  12-G2&  12-G3&  12-G4  \\

\multicolumn{12}{c}{} \\
\multicolumn{12}{c}{} \\
\multicolumn{12}{c}{\emph{NGC 2023 South}} \\

11.0	&		&		&		&		&		&		&		&		&		&		&		\\
11.2	&	0.52	&		&		&		&		&		&		&		&		&		&		\\
11-G1	&	\textcolor{red}{\textbf{0.97}}	&	0.50&		&		&		&		&		&		&		&		&		\\
11-G2	&	0.66	&	0.86	&	0.60&		&		&		&		&		&		&		&		\\
11-G3	&	0.81	&	0.60&	0.80&	0.70&		&		&		&		&		&		&		\\
11-G4	&	0.47	&	\textcolor{red}{\textbf{0.99}}	&	0.44	&	0.81	&	0.53	&		&		&		&		&		&		\\
11-G5	&	0.55	&	\textcolor{red}{\textbf{0.99}}	&	0.53	&	0.87	&	0.63	&	\textcolor{red}{\textbf{0.97}}	&		&		&		&		&		\\
12-G1	&	0.66	&	0.87	&	0.63	&	0.83	&	0.70&	0.83	&	0.87&		&		&		&		\\
12-G2	&	0.68	&	0.61	&	0.65	&	0.72	&	0.77	&	0.56	&	0.63	&	0.55	&		&		&		\\
12-G3	&	0.84	&	0.53	&	0.83	&	0.70&	\textcolor{red}{\textbf{0.92}}	&	0.46	&	0.57	&	0.61	&	0.82	&		&		\\
12-G4	&	0.55	&	\textcolor{red}{\textbf{0.95}}	&	0.52	&	0.86&	0.60&	\textcolor{red}{\textbf{0.93}}	&	\textcolor{red}{\textbf{0.96}}	&	0.87	&	0.64	&	0.53	&		\\
\midrule
	&   11.0&   11.2&  11-G1&  11-G2&  11-G3&   11-G4&  11-G5&   12-G1&  12-G2&  12-G3&  12-G4  \\

\multicolumn{12}{c}{} \\
\multicolumn{12}{c}{} \\
\multicolumn{12}{c}{\emph{NGC 2023 North}} \\

11.0	&		&		&		&		&		&		&		&		&		&		&		\\
11.2	&	0.61	&		&		&		&		&		&		&		&		&		&		\\
11-G1	&	\textcolor{red}{\textbf{0.99}}	&	0.63	&		&		&		&		&		&		&		&		&		\\
11-G2	&	0.76	&	0.89	&	0.75	&		&		&		&		&		&		&		&		\\
11-G3	&	0.66	&	\textcolor{red}{\textbf{0.93}}	&	0.68	&	\textcolor{red}{\textbf{0.91}}	&		&		&		&		&		&		&		\\
11-G4	&	0.56	&	\textcolor{red}{\textbf{0.99}}	&	0.57	&	0.85	&	\textcolor{red}{\textbf{0.90}}&		&		&		&		&		&		\\
11-G5	&	0.68	&	\textcolor{red}{\textbf{0.98}}&	0.70&	\textcolor{red}{\textbf{0.92}}	&	\textcolor{red}{\textbf{0.93}}	&	\textcolor{red}{\textbf{0.94}}	&		&		&		&		&		\\
12-G1	&	0.64	&	0.76	&	0.64	&	0.78	&	0.77	&	0.75	&	0.76	&		&		&		&		\\
12-G2	&	0.64	&	0.64	&	0.65	&	0.69	&	0.75	&	0.60&	0.68	&	0.43	&		&		&		\\
12-G3	&	0.87	&	0.76	&	0.88&	0.88	&	0.83	&	0.70&	0.82	&	0.64	&	0.78	&		&		\\
12-G4	&	0.77	&	0.76	&	0.76	&	0.83	&	0.79	&	0.71	&	0.83	&	0.74	&	0.70&	0.80&		\\
\midrule
	&   11.0&   11.2&  11-G1&  11-G2&  11-G3&   11-G4&  11-G5&   12-G1&  12-G2&  12-G3&  12-G4  \\
\end{tabular}
\end{center}
\end{table*}

\subsection{The 11.0 and 11.2 \mt bands: assignments}
\label{sec:charge}

Since early on, astronomical observations of extended sources have revealed that the major PAH bands exhibit spatially different behaviour. In particular, the 8.6 and 11.0 \mt bands peak closer to the exciting star than the 3.3 and 11.2 \mt bands, which peak further away (e.g. \citealt{joblin1996,sloan1999}). The behaviour has been attributed to the PAH charge state, as laboratory experiments have shown that the 8.6 and 11.0 \mt emission are dominated by cations, and the 11.2 \mt emission by neutral PAHs \citep{allamandola1999,hudgins1999}. A comparison of astronomical spectra to laboratory and theoretically calculated spectra by \cite{hony2001} reinforced the assignment of the 11.0 and 11.2 \mt bands to solo \choop bending in cations and neutrals, respectively. More recently, computed spectra of (very) large PAHs by \cite{bauschlicher2008} and \cite{ricca2012} showed that the solo \choop emission from PAH neutrals becomes blueshifted upon ionization, supporting the same assignment. Using blind signal separation, \cite{rosenberg2011} also identified the 11.0 \mt band as cationic and the 11.2 \mt band as neutral.

Recently, however, there have been suggestions of further complexity in the assignments of the 11.0 and 11.2 \mt bands, which we address here. For one, \cite{candian2015} showed that \textit{neutral} acenes produce emission near 11.0 \m. At present it is not clear how significant their contribution will be to the astronomical 11.0 \mt emission band, as they constitute a small set of the PAH family. The NASA Ames PAH IR Spectroscopic Database\footnote{\url{http://www.astrochem.org/pahdb/}} \citep{bauschlicher2010,boersma2014_amesdb}, hereafter referred to as PAHdb, has few included acenes at this time. Using PAHdb, \cite{boersma2013} showed that \textit{cationic} nitrogen-substituted PAHs, or PANHs, were required to fit the 11.0 \mt astronomical emission in NGC 7023. Another possibility is that [SiPAH]$^+$ complexes may contribute to the 11.0 \mt band, as shown by quantum chemical calculations by \cite{joalland2009}. The resulting IR emission intensities of such complexes are expected to be similar to those of pure PAH cations. To confirm this assignment further laboratory and theoretical work are required \citep{joalland2009}. Regarding the 11.2 \mt emission, it has been proposed that its red wing (out to 11.4-11.6 \m) is due to the emission from very small grains (VSGs; \citealt{berne2007,rosenberg2011}). If VSG abundances are sufficiently high, they may influence the peak position of the 11.2 \mt complex.

Apart from neutral acenes, all assignments of the 11.0 \mt band point towards a cationic carrier. Similarly, apart from VSGs, all assignments of the 11.2 \mt band point towards neutral PAHs. We adopt such charge assignments, leading to the following conclusions:

\begin{enumerate} \itemsep1pt \parskip0pt \parsep0pt

\item The 11 \mt complex:
\begin{enumerate} \itemsep1pt \parskip0pt \parsep0pt
	\item[] 11-G1 --- dominated by cations
	\item[] 11-G2 --- mixed charge
	\item[] 11-G3 --- mixed charge
	\item[] 11-G4 --- dominated by neutrals
	\item[] 11-G5 --- dominated by neutrals
\end{enumerate}

\item The 12.7 \mt complex:
\begin{enumerate} \itemsep1pt \parskip0pt \parsep0pt
	\item[] 12-G1 --- mixed charge
	\item[] 12-G2 --- mixed charge
	\item[] 12-G3 --- dominated by cations
	\item[] 12-G4 --- dominated by neutrals
\end{enumerate}

\end{enumerate}

\subsection{Neutral emission at 11.0 \m}

\begin{figure}
	\centering
	\includegraphics[width=\linewidth]{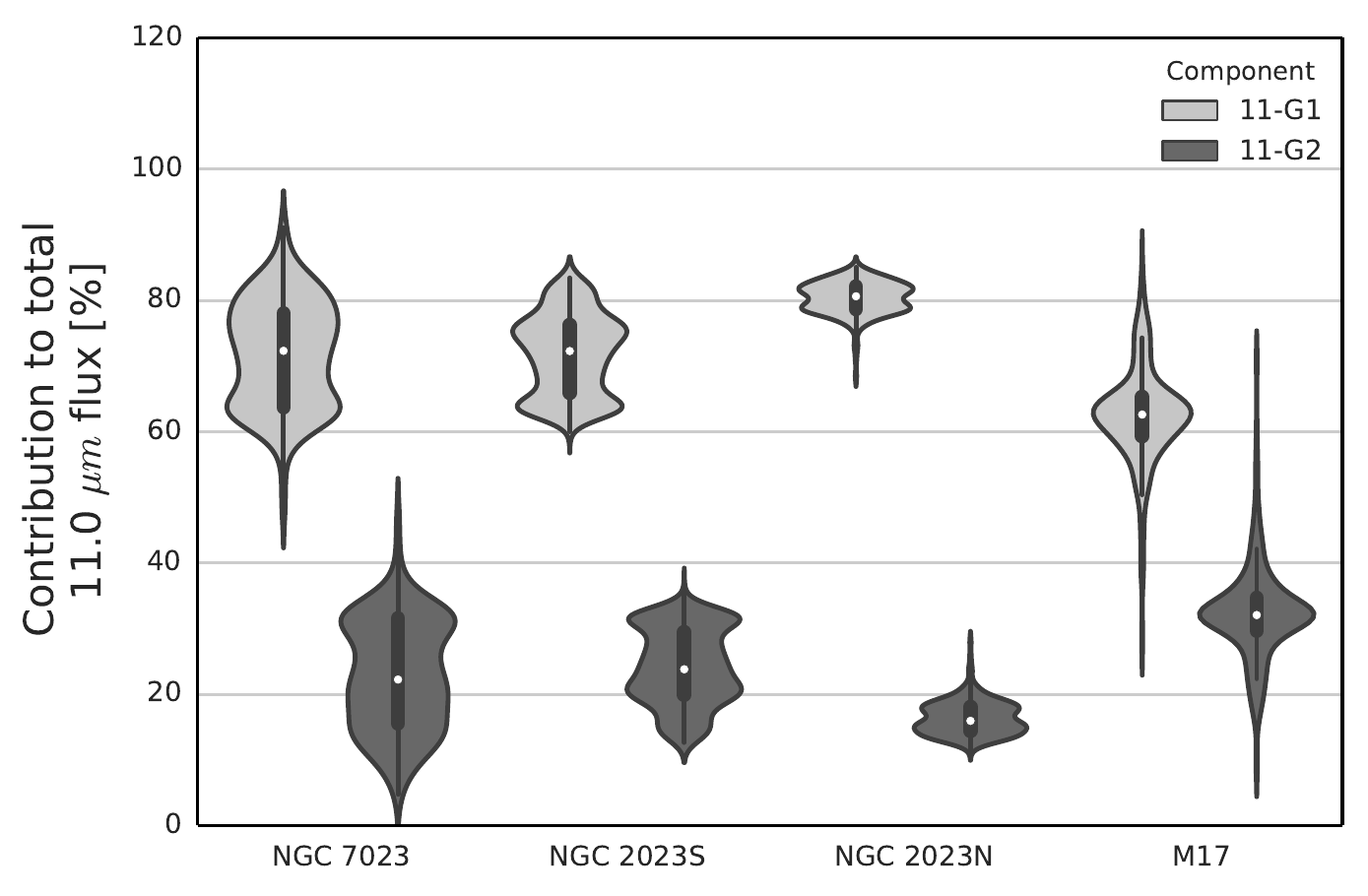}
	\caption{Breakdown of the fractional contributions to the traditional 11.0 \mt flux from the 11-G1 and 11-G2 components for each object. The violin shape is a histogram of all pixels in each data cube, shown per object. The violin width corresponds to the number of pixels contributing at a given percent level. The total histogram is fitted with a kernel density estimate (which leads to the violin shape). The mean values are indicated by the white dots and the standard deviation by the thick dark line. The thin dark line (running vertically through each shape) represents the total data range.
	}
	\label{fig:stats}
\end{figure}

The mixed behaviour of the 11-G2 component indicates that the traditional 11.0 \mt band is not purely cationic. The 11-G1 component however dominates the 11.0 \mt flux, meaning that the emission is still primarily cationic. We estimate the contributions of these two components to the traditional 11.0 \mt emission in Fig.~\ref{fig:stats}. The 11-G1 component in general carries 60-80\% of the flux: $72\pm8$\% (NGC 7023), $72\pm6$\% (NGC 2023 South), $80\pm3$\% (NGC 2023 North) and $62\pm7$\% (M17). The 11-G2 component generally is the complement to these values: $23\pm8$\% (NGC 7023), $24\pm6$\% (NGC 2023 South), $16\pm3$\% (NGC 2023 North) and $32\pm7$\% (M17). In some instances there is a minor contribution (less than 5\%) from either the 11-G3 or 11-G4 component.

We use the structural similarity indices (Table~\ref{tables:ssim7023}) to quantify the mixed behaviour of the 11-G2 component. To do so, we must make the assumption here that the traditional 11.2 \mt component traces neutral PAHs and the 11-G1 component traces pure cations. We can reach the same conclusions if we instead use the traditional 11.0 \mt band as a tracer of pure cations, though with greater uncertainty (since the 11-G2 component is a minor contributor to the 11.0 \mt emission).  The similarity indices show that in NGC 7023, the 11-G2 component is equally similar to the 11.2 \mt emission (SSIM = 0.73) as it is to the 11-G1 emission (SSIM = 0.74). In NGC 2023 South and North, however, the 11-G2 component is more similar to the 11.2 \mt band than the 11-G1 emission (SSIM = 0.86 versus 0.60 in the southern map and SSIM = 0.89 versus 0.75 in the northern map). This implies that at minimum 50\% of the flux of the 11-G2 component is from neutral PAHs. In turn, this then implies that \textit{the traditional 11.0 \mt band is roughly $8-16$\% neutral and $78-88$\% cationic, with uncertainties approaching ten percentage points.}

\bigskip

Recently, \citet{candian2015} reported that the solo \choop mode of neutral acenes falls near 11.03 \mt and argues for a possible contribution from neutral acenes to the 11.0 \mt band. The results reported here, that there is a small neutral contribution to the 11.0 \mt band in our sample, provide the first observational evidence in support of a partial origin in a neutral carrier.

\bigskip

\cite{rosenberg2011} applied blind signal separation to NGC 7023 and identified three basis vectors: PAH$^0$, PAH$^+$ and VSGs. One interesting result of their analysis is that the PAH$^0$ signal, which is associated with neutral PAHs, showed a local emission peak at 11.0 \m. The authors suggest this is an artifact of the applied method, and may be compensated for by a local minimum in their VSG signal. However, since we have identified a component of the traditional 11.0 \mt band that is linked to neutral PAHs, we suggest that it may be due to the non-cationic component that we have found.

In the traditional assignment of the 11.0 and 11.2 \mt bands, i.e. originating from solo \choop modes in cations and neutrals, respectively, the 11.0/11.2 ratio traces the PAH charge fraction \citep{boersma2012}. To examine this, one must determine the intrinsic 11.0/11.2 flux ratio of a PAH after a single photon excitation event. The authors chose circumcoronene as being representative of a typical PAH, which then relates the observed 11.0/11.2 flux ratio to an implied neutral-to-cation fraction. The neutral fraction was shown to decrease in Orion from 80\% to 65\% when moving away from the star, before returning to 80\% at further distances. The unexpected diminution of the neutral fraction with increasing distance may reflect dehydrogenation, which would affect the measured 11.0/11.2 \mt ratio, or it may indicate that circumcoronene is not a reasonable proxy for the total PAH population \citep{boersma2012}. Our results show that neutral PAHs contribute to the 11.0 \mt band, at approximately the 8-16\% level. This may be an additional contributing factor to the unusual 11.0/11.2 behaviour observed by \cite{boersma2012}.

Mean spectra of species in PAHdb were presented by \cite{boersma2013}, binned by size and charge state (their Fig. 9). There does not appear to be neutral emission at 11.0 \m. This may reflect biases or limitations of the database (see \citealt{bauschlicher2010,boersma2011}). The neutral emission we deduce to exist at 11.0 \mt is a small fraction of the cationic emission at 11.0 \m, and thus it may be hidden (if present in the PAHdb spectra) by averaging over all species.

\subsection{Profile asymmetries}

\subsubsection{The 11 \mt emission}

The 11.2 \mt band displays two asymmetries: a prominent red wing in the range 11.4-11.6 \mt and a narrow peak near 11.20 \mt that appears only when sufficiently close to the illuminating source (c.f. Fig.~\ref{fig:raw}). We exclude the 11.0 \mt band from consideration here as it is understood to be a separate band.

In our decomposition, the 11-G3, 11-G4 and 11-G5 components all emit significantly at 11.20 \mt (Fig.~\ref{fig:112_3pos}). However, the 11-G3 component is much narrower than the others and it is located at exactly 11.20 \m, meaning that it might provide some clues about the origin of the peak asymmetry. With the caveats in mind, the results show that the spatial distribution of the 11-G3 component is a mixture of the traditional 11.0 and 11.2 \mt emission. The 11-G3 component has almost no spectroscopic overlap with the traditional 11.0 \mt emission, and the 11-G3 component is strongly blended at 11.2 \mt with the neutral-carrying 11-G4 and 11-G5 bands, which dominate the 11.2 \mt flux. This implies that \textit{we identify a cationic contribution to the 11.2 \mt emission.}

Furthermore, the strength of the 11-G1 and 11-G3 components both increase when approaching the star (as shown in Fig.~\ref{fig:112_3pos}). This is complemented by a transition of the 11.2 \mt peak position from class A(B) (near 11.24 \m) to class A (near 11.20 \m) as reported by \cite{boersma2012,boersma2013} (see \citealt{peeters2002,vandiedenhoven2004} for classification details). Since the 11-G3 component is coincident with the nominal class A position, it suggests that it is the relative intensity of the 11-G3 component to that of 11.2 \mt emission peak that determines the PAH class. Knowing also that the 11-G3 component has a cationic contribution, this implies that the PAH classification of the 11.2 \mt band is partially moderated by the relative fraction of emission from PAH cations to neutrals. This effect is likely in addition to that reported by \cite{candian2015}, who studied neutral PAHs and identified that the class variations, from A to A(B), result from changes in the distribution of PAH masses.

\bigskip

During the class transition from A(B) to A, the flux in the red wing decreases significantly in NGC 7023 (c.f. Fig.~\ref{fig:raw}), and less so in NGC 2023 South. The spectral asymmetry due to the variable red wing is well known in the literature \citep{hudgins1999,hony2001,vandiedenhoven2004}. Many PAH bands display a red wing due to anharmonicity, but the magnitude of variations seen in the observations cannot be reconciled with the expected degree of asymmetry from this effect \citep{vandiedenhoven2004}. Possible explanations have been presented in the literature but its origin has not yet been established. Suggestions include emission from VSGs \citep{rapacioli2006,berne2007,rosenberg2011}, PAH clusters \citep{boersma2014}, PAH anions \citep{bauschlicher2009}, low-mass PAHs \citep{candian2015} and superhydrogenated PAHs \citep{knorke2009,boersma2014}.

\cite{rosenberg2011} found a spatial separation between neutral PAHs and very small grains in NGC 7023 based on blind signal separation. In our decomposition, the flux of the red wing is carried by the 11-G4 component, while the ``symmetric" 11.2 \mt emission is carried primarily by the 11-G5 component. As introduced in Section~\ref{sec:results_maps}, the spectral maps of the 11-G4 and 11-G5 components are similar but have subtle differences. We observe in NGC 7023 and 2023 South that the 11-G4 (red wing) emission is less extended than the 11-G5 emission, despite peaking in the same map position. The discrepancy with the results of \cite{rosenberg2011} is likely a consequence of the large width of our 11-G4 component, which has significant contributions at 11.1 \mt and greater.

\bigskip

We conclude that the entire 11 \mt complex traces the following (not necessarily unique) populations: at 11.0 \m, PAH cations and a small fraction of neutrals; and at 11.2 \m, PAH neutrals and a small fraction of PAH cations. With our decomposition we are unable to deduce the carrier of the red wing.

\subsubsection{The 12.7 \mt emission}

The 12.7 \mt band is quite asymmetric, exhibiting a blue-shaded wing \citep{hony2001}. Our decomposition accounts for the blue wing primarily through the flux of the 12-G1 component, which is broad and centered at 12.55 \m. In NGC 7023, the 12-G1 component is very extended, easily encompassing the regions containing strong 11.0 and 11.2 \mt emission. This suggests that the broad component, as defined here, has no charge preference. In NGC 2023 South, it is also quite extended, though to a lesser extent than in NGC 7023. The structural similarity indices show that the map of the 12-G1 component is equal parts 11.0 and 11.2 \mt emission in NGC 7023. The 12-G1 component is slightly more similar to the 11.2 \mt emission than the 11.0 \mt emission in NGC 2023 South. Similar results to NGC 2023 South are found in the northern map, suggesting only a weak dependence on charge. This likely originates in the broadness of the component and reflects that the proposed decomposition does not completely disentangle all PAH sub-populations.

\cite{bauschlicher2008,bauschlicher2009} used density functional theory (DFT) to compute the absorption spectra of large symmetric PAHs and large irregular PAHs. Near 12.7 \m, these authors found that the mean emission from PAH cations is blueshifted from the neutral position. In our decomposition, we find the same relationship: the 12-G3 component at 12.72 \mt is cationic, while the 12-G4 component at 12.78 \mt is neutral. Hence, the intrinsic spectra and observations are consistent with each other.

\cite{boersma2013} decomposed the PAH emission in NGC 7023 with PAHdb. In fitting the 12.7 \mt emission the authors find that, in the dense region, neutrals are responsible for the majority of the total intensity. In the diffuse region(s), the cations instead carry most of the intensity\footnote{The ``diffuse region" of \cite{boersma2013} refers to the region between the star and PDR front. The ``dense region" refers to the region beyond that.}. We observe that the ratio of the cationic 12-G3 emission to the neutral 12-G4 emission is higher in the diffuse region than the dense region, consistent with their result. \cite{boersma2013} also found that the dense region was characterized by small PAHs (those with fewer than 50 carbon atoms) and the diffuse region was dominated by large PAHs (those with more than 50 carbon atoms). Within this framework, the 12-G3 component then originates in large PAHs preferentially. The 12-G4 component corresponds to emission that was comparable between small and large PAHs in the PAHdb fit \citep{boersma2013}, suggesting an equal mixture of sizes.

It has been noted in the literature that some PAH correlation plots involving the 12.7 \mt band display a bifurcation \citep{boersma2014,stock2014}. As our results show that the profile of the 12.7 \mt band can be strongly dependent on ionization, perhaps the bifurcated correlations simply trace the two different charge states of the 12.7 \mt emission.

\subsection{The 12.7/11.2 intensity ratio}
\label{subsec:127_over_112}

\begin{figure*}
	\centering
	\includegraphics[width=\linewidth]{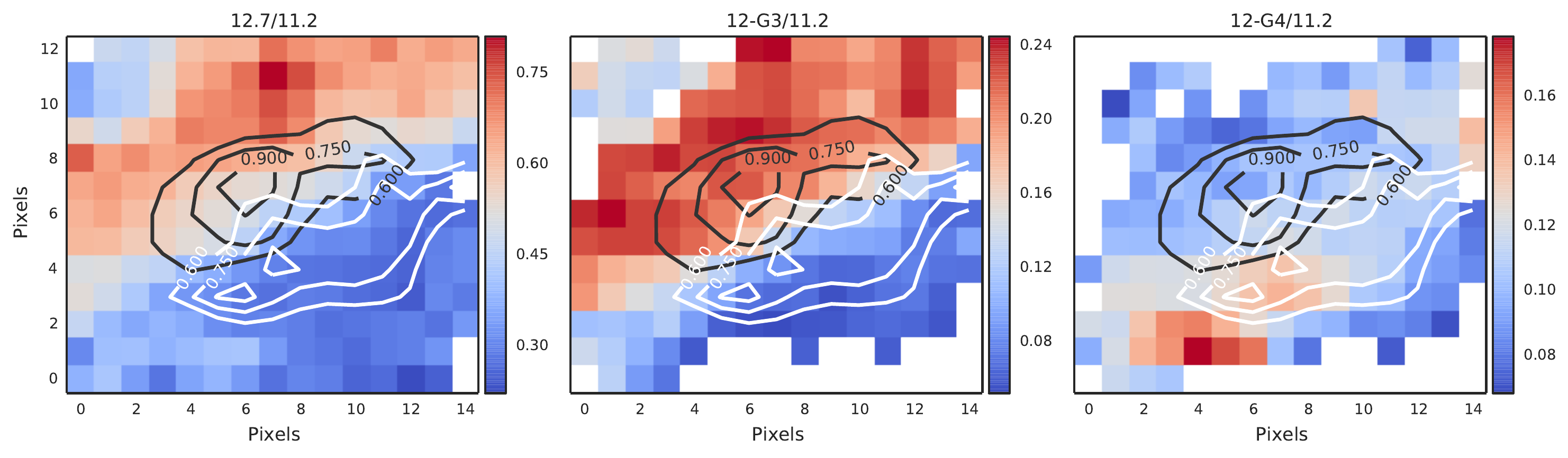}
	\caption{The 12.7/11.2 ratio in NGC 7023 is compared to the 12-G3/11.2 and 12-G4/11.2 ratios (left, middle and right panels, respectively). The black contours are constructed from the traditional 11.0 \mt emission and the white contours from the 11.2 \mt emission.
	}
	\label{fig:map_7023_127}
\end{figure*}

The 12.7/11.2 ratio has long been understood as probing hydrogen adjacency, as the 11.2 \mt band is associated with solo \choop bending modes and the 12.7 \mt band with duo and trio \choop bending modes \citep{hony2001,bauschlicher2008}. The astronomical range of 12.7/11.2 intensity ratios has been shown to be consistent with computed spectra of irregular PAHs, which have solos, duos, trios and quartets \citep{ricca2012,bauschlicher2009}. PAHs with only solos and duos, such as those from the coronene family, lead to 12.7/11.2 ratios that lie at the low end of the astronomical range \citep{ricca2012}. In addition, this ratio can be further enhanced due to the coupling between the \choop modes and the C-C ring deformation mode in (elongated) armchair PAHs contributing to the 12.7 \mt emission \citep{candian2015}.

Through the use of PAHdb, \cite{boersma2015} showed that the spatial variation of the 12.7/11.2 ratio in NGC 7023 seems to primarily trace ionization rather than edge structure. Based on their database fits, the authors observed that the ionization fraction increases by 200\% across NGC 7023, whereas the hydrogen adjacency only drops by 25\%. Since we have presented a way of generally disentangling ions from neutrals in the 11 and 12.7 \mt complexes, we can probe the dependence of the 11.2/12.7 ratio on charge and molecular edge structure.\footnote{To probe charge and hydrogen adjacency, we first use our decompositions to isolate the neutral and cationic contribution to the 11 and 12.7 \mt complexes. To do this, we use the charge breakdown adopted in Section~\ref{sec:charge}. For the mixed charge bands (e.g. 11-G2, 11-G3), we use the spatial maps and structural similarity indices to discern if one charge state appears to be dominant. The 11-G2 and 12-G2 components are found to be on average equally similar to the cationic- and neutral-dominated bands. Thus, we assume half of their flux contributes to the total cationic emission, and half to the neutral emission. For the other mixed charge bands (11-G3, 12-G1), we tested a series of different charge fractions (25\%, 50\%, or 75\% cationic). The fluxes of these components relative to the other fluxes in the calculation are sufficiently low that they do not affect our conclusions. Since the 11 \mt emission is thought to trace solo \choop structures, and the 12.7 \mt emission duo and trio \choop structures, we finally measure the solo/duo+trio ratio for cations, and separately for neutrals.} We found that the ionization fraction (defined as the flux from cations divided by the flux from both cations and neutrals) spans a factor of of 3.1 $\pm$ 1.4 across the NGC 7023 map. Although the uncertainty is large, the derived ionization fraction is very similar to the 200\% increase (i.e. a factor of 3) determined by \cite{boersma2015}. To study hydrogen adjacency, we measured the 12.7/11.2 ratio separately for cations and neutrals. For cations, hydrogen adjacency varies across the map by approximately 30\%, while for neutrals it varies by about 10\%. These results are generally consistent with the 25\% value presented by \cite{boersma2015}. We also applied this analysis to NGC 2023 South and found that the ionization fraction spans a range of 1.9 $\pm$ 0.8 (or a 100\% increase across the map). The hydrogen adjacency varied by approximately 25\% for cations, but only 7 or 8\% for neutrals. This analysis shows that the 12.7/11.2 ratio depends largely on the ionization fraction and to a lesser extent on molecular edge structure. However, when charge state is taken into account (by only considering neutral PAHs or only PAH ions), we can trace molecular edge structure of the PAH population.

The spatial morphology of the 12.7/11.2 emission was compared against the spatial distribution of the 6.2/($6.2+11.2$ \m) fraction and the PAHdb-derived fractional emission in PAH cations \citep{boersma2014}. The latter two both display smooth gradations across the PDR, whereas the 12.7/11.2 map shows pockets of enhanced ratios. The authors interpreted this as a consequence of the mixed-charge behaviour of the 12.7 \mt band. We re-examine this relationship in Fig.~\ref{fig:map_7023_127}. We now include two additional maps: the 12-G3/11.2 and 12-G4/11.2 ratios.  We find a stark increase in the contrast between the diffuse and dense regions when examining 12-G3/11.2 instead of 12.7/11.2. This originates in the fact that, to first order, the neutral dependence is removed. The 12-G4/11.2 ratio shows little emission in the diffuse region and relatively little variation across the region of peak 11.2 \mt emission. One maximum is observed in the dense region. Clearly the 12-G3/11.2 and 12-G4/11.2 ratios are tracing different PAH populations. Taking these effects into account, it is apparent that the 12.7/11.2 ratio probes both ionization and molecule structure, which both depend on the local physical conditions (c.f. \citealt{boersma2015}).

\begin{figure*}
	\centering
	\includegraphics[width=\linewidth]{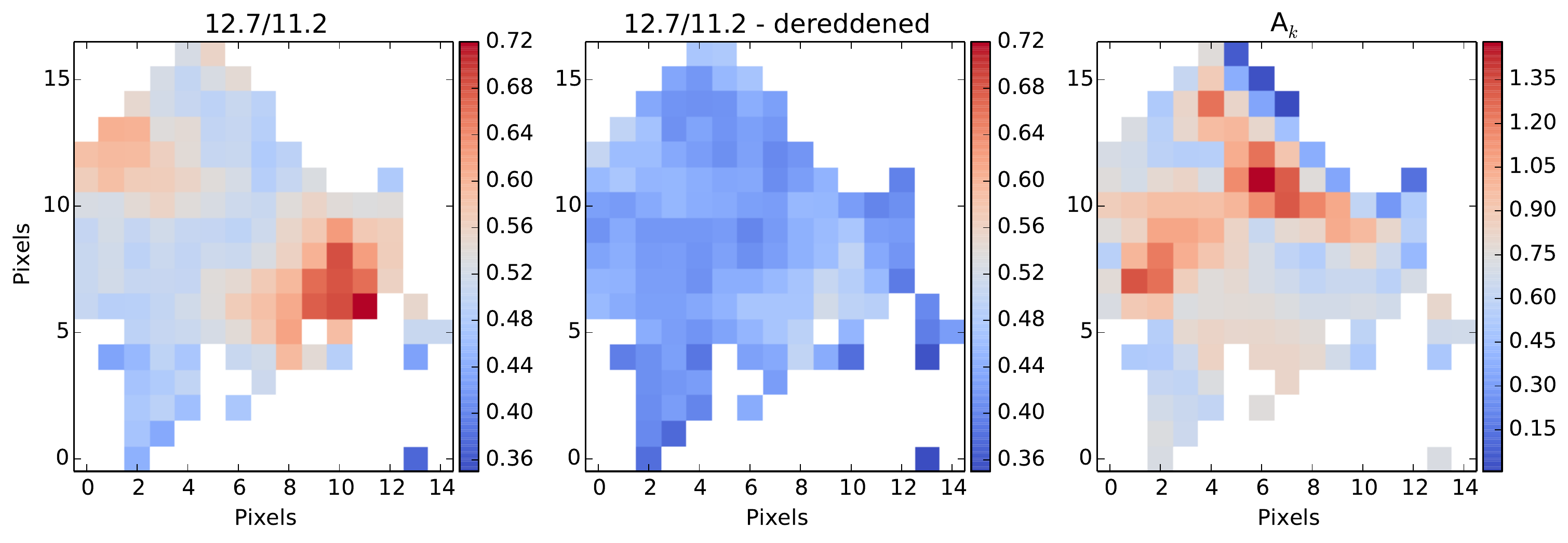}
	\caption{The effect of extinction on the 12.7/11.2 ratio in M17 is examined. Left panel: the 12.7/11.2 ratio with no correction. Middle panel: the extinction corrected 12.7/11.2 ratio. Right panel: the extinction map used for the correction.
	}
	\label{fig:extinction}
\end{figure*}

It's worth noting that the 12.7/11.2 ratio can be greatly affected by extinction. To illustrate this, we quantify this effect by comparing the 12.7/11.2 ratio before and after correcting for extinction in M17 (Fig.~\ref{fig:extinction}) and two \HII~regions from \cite{hony2001}. Using the interstellar extinction curves of \cite{chiar2006}, we corrected for extinction. In M17, the 12.7/11.2 ratio decreases from a range of 0.37-0.72 (before correction) to 0.35-0.52 (after correction). Some pixels exhibit a greater than 30\% decrease in the 12.7/11.2 ratio. We also examined two \HII~regions from the sample of \cite{hony2001}, using the extinction measurements of \cite{martin-hernandez2002}: IRAS 15384-5348 (A$_k=1.3$) and IRAS 18317-0757 (A$_k=2.0$). The latter source exhibits the largest 12.7/11.2 ratio in the study of \cite{hony2001} before extinction correction (1.49), but this value is near unity after correction (1.03). Extinction correction thus significantly reduces the large range in 12.7/11.2 ratios observed in \HII~regions.

With regards to the 12.7/11.2 ratio in NGC 7023, only three pixels in the map meeting the signal-to-noise criterion are affected by extinction to any substantial degree (approximately 10-15\% flux difference in the 12.7 \mt band). These pixels are on the lower edge of the map and do not affect the map or our conclusions.

\section{Conclusions}
\label{sec:conclusion}

We have examined the spatial and spectral behavior of the 11 and 12.7 \mt PAH emission complexes in high-resolution \textit{Spitzer}/IRS maps of NGC 7023, NGC 2023 South and North and M17. We have introduced a five-component Gaussian decomposition of the 11 \mt emission and a four-component decomposition of the 12.7 \mt emission. At 11 \m, two components are centered near 11.0 \m, one of which is relatively broad and the other relatively narrow. A narrow, weak Gaussian component is located near the peak of the 11.2 \mt complex, slightly offset towards the blue. A strong band carries most of the peak 11.2 \mt emission, and a final broad emission component is responsible for the red wing of the 11.2 \mt band. The 12.7 \mt decomposition consists of one broad feature on the blue wing, two components near 12.75 \m, and a narrow feature that appears on the blue wing in some positions.

We investigated the spatial distributions of these components in our spectral maps and quantified their similarities the structural similarity index, in addition to flux correlation plots. We have arrived at the following conclusions:

\begin{enumerate}

\item The traditional 11.0 \mt emission band has a contribution from neutral PAHs. We identify a broad cationic feature at 11.00 \mt and a weaker, narrower mixed-charge state band at 11.00 \m. In total, 78-88\% of the traditional 11.0 \mt flux is carried by cations, and 8-16\% of the flux is carried by neutral PAHs. This may be supporting evidence for the contribution of neutral acenes to the 11.0 \mt band \citep{candian2015}.

\item The traditional broad 11.2 \mt emission band has a small cationic contribution at 11.20 \m. The relative strength of this cationic feature to the broad 11.2 \mt emission determines the overall peak position of the 11.2 \mt complex. This implies that the PAH classification of the 11.2 \mt emission is partially determined by the fraction of PAH cations to neutrals, in addition to the varying distribution of PAH masses reported by \cite{candian2015}.

\item The variable peak position of the 12.7 \mt complex can be explained by the relative strengths of two competing Gaussian components at 12.72 and 12.78 \m. The features are spectroscopically blended, yet they distinctly trace cations and neutrals, respectively, in the spectral maps.

\item The component responsible for the bulk of the blue-shaded wing of the 12.7 \mt band appears to be only weakly dependent on charge. This may indicate that the proposed decomposition does not completely disentangle all PAH sub-populations.

\item The observed contribution of both cation and neutral PAHs to the 12.7 \mt band supports the use of the 12.7/11.2 intensity ratio as a charge proxy \citep{boersma2015}. However, after accounting for PAH charge, structural variations are still probed by the 12.7/11.2 ratio.

\end{enumerate}

These results illustrate the power of spectral maps for understanding the complicated spectral profiles of PAHs, wherein blended spectral components can be understood as independent spatial components. The next step is to apply this technique to other high-resolution spectral maps as well as integrated spectra of individual objects. This will help improve the quantification of the non-cationic emission at 11.0 \mt and understand the behaviour of the broad blue 12.7 \mt wing, which appears to be insensitive to charge based on the applied decomposition. This work and the interpretation of the PAH emission bands will strongly benefit from the heightened spectral resolution and sensitivity of the forthcoming James Webb Space Telescope mission.

\section*{Acknowledgments}

The authors acknowledge support from an NSERC discovery grant, NSERC acceleration grant and ERA. MJS acknowledges support from a QEII-GSST scholarship. The IRS was a collaborative venture between Cornell University and Ball Aerospace Corporation funded by NASA through the Jet Propulsion Laboratory and Ames Research Center \citep{houck2004}. This research has made use of NASA's Astrophysics Data System Bibliographic Services, and the SIMBAD database, operated at CDS, Strasbourg, France. This work has also made use of the Matplotlib Python plotting library \citep{hunter2007} and the Seaborn Python visualization library\footnote{\url{http://dx.doi.org/10.5281/zenodo.19108}}.

\appendix
\renewcommand{\thefigure}{A\arabic{figure}}
\setcounter{figure}{0}

\section{Spatial Maps of NGC 2023 North and M17}

\begin{figure*}
	\centering
\includegraphics[width=0.9\linewidth, trim={0cm 9.5cm 0cm 0cm}, clip]{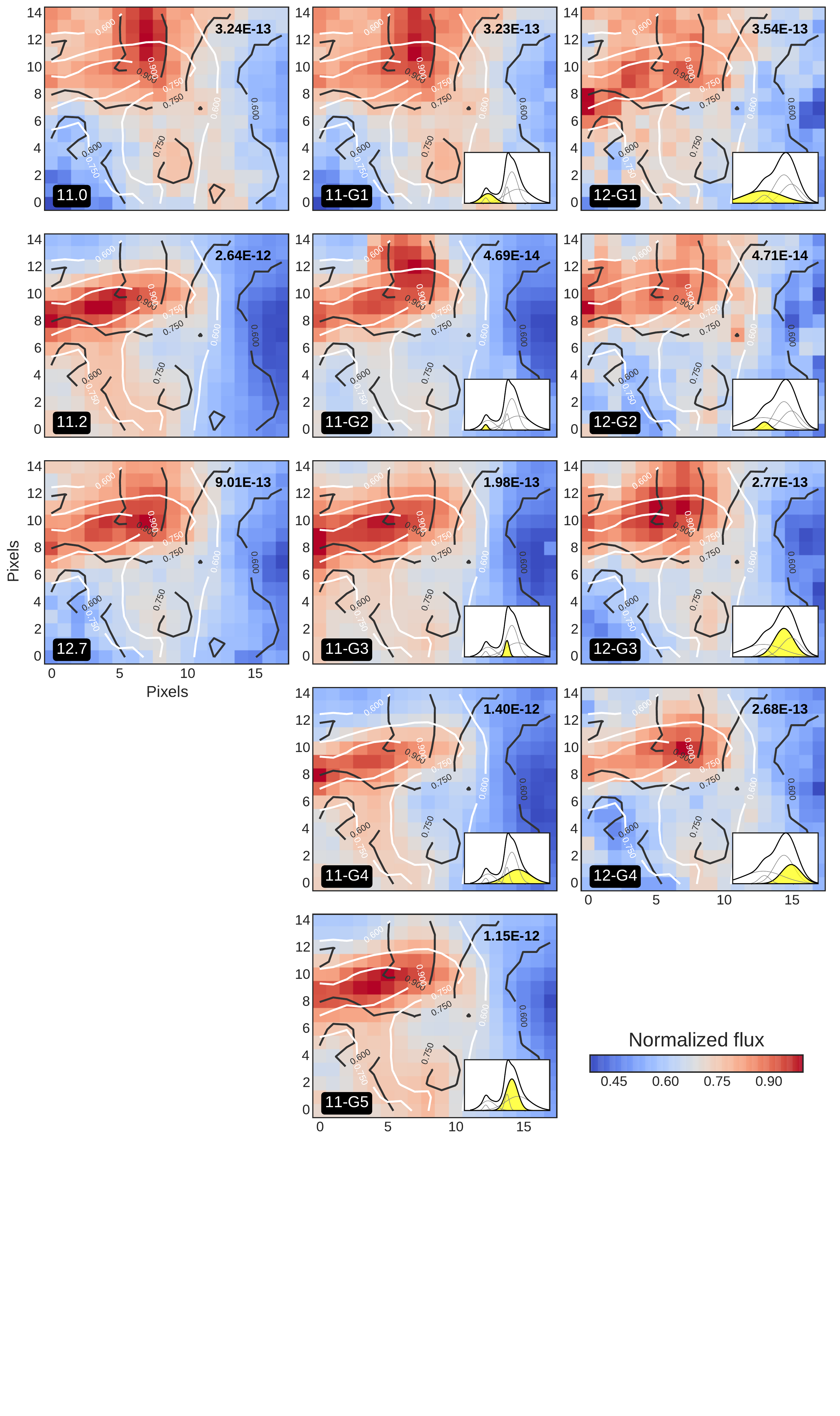}
	\caption{PAH emission in NGC 2023 North. The conventions are the same as those used in Fig.~\ref{fig:map_7023}.
	}
	\label{fig:map_2023N}
\end{figure*}

\begin{figure*}
	\centering
\includegraphics[scale=0.37, trim={0cm 0cm 0cm 0cm}, clip]{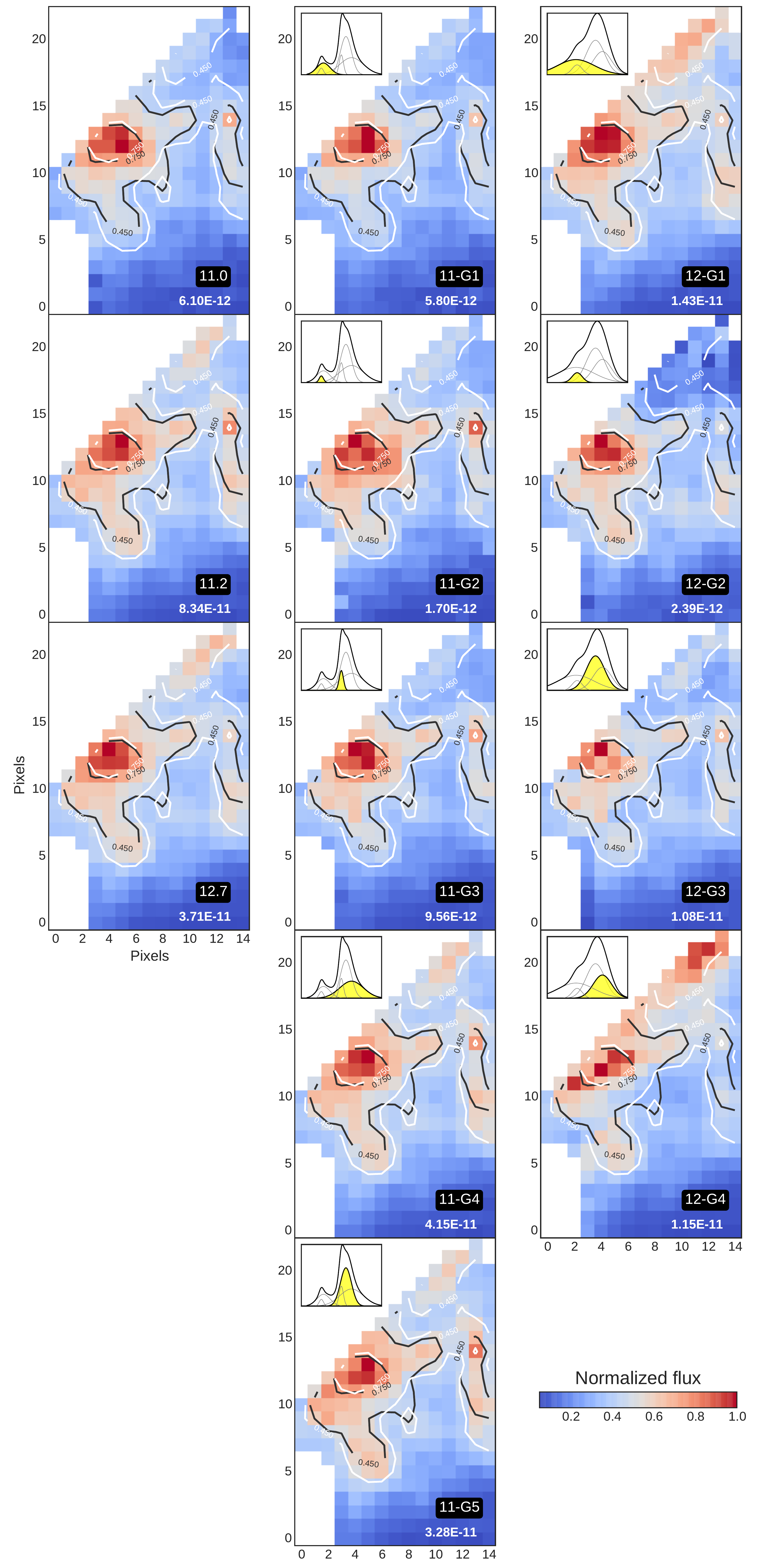}
	\caption{PAH emission in M17. The conventions are the same as those used in Fig.~\ref{fig:map_7023}.
	}
	\label{fig:map_M17}
\end{figure*}

The maps of NGC 2023 North and M17 are presented in Figs.~\ref{fig:map_2023N} and~\ref{fig:map_M17}, respectively. These are discussed in the main text in Sections~\ref{sec:11decomp} and~\ref{sec:12decomp}. Here we give a brief summary.

In general, NGC 2023 North displays a much more complex morphology than NGC 2023 South or NGC 7023. The peak PAH emission appears to be oriented either vertically in this map (e.g. 11.0 \m, 11-G1 emission) or horizontally (e.g. 11.2 \m, 11-G3, 11-G4, 11-G5, 12-G4 emission), with significant overlap. However, the similarities between bands found in NGC 2023 North are generally consistent with those found for NGC 2023 South and NGC 7023 (see main text). Likewise, M17 is a complicated environment for analysis. In Fig.~\ref{fig:map_M17}, we observe that the peak 11.0 and 11.2 \mt emission regions are nearly coincident. However, the key distinguishing feature seems to be the M17 ``spur" emission in the upper right corner of the map, which is stronger in the 11.2 \mt band. Based only on this distinguishing feature, key results from studying NGC 7023 and NGC 2023 South are also present here. The most robust trends---that the 11.0 \mt and 12-G3 emission are well-matched, as are the 11.2 \mt and 12-G4 emission---are both observed in M17 when examining the spur, despite its complexity.

\section{Flux correlation plots}

\begin{figure*}
\begin{center}
    \subfigure{
	\includegraphics[width=0.8\linewidth]{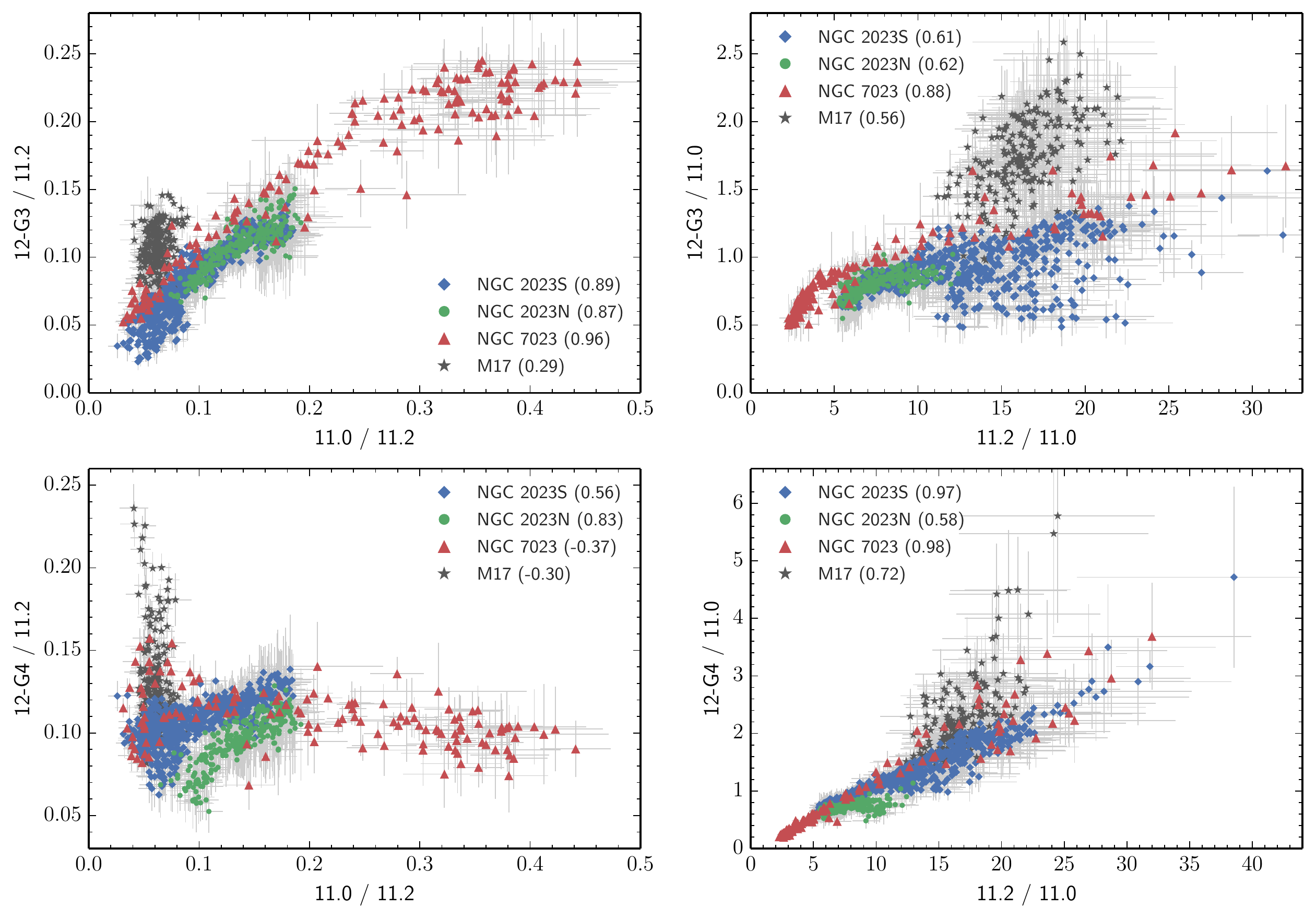}
    } \\
    \subfigure{
	\includegraphics[width=0.8\linewidth]{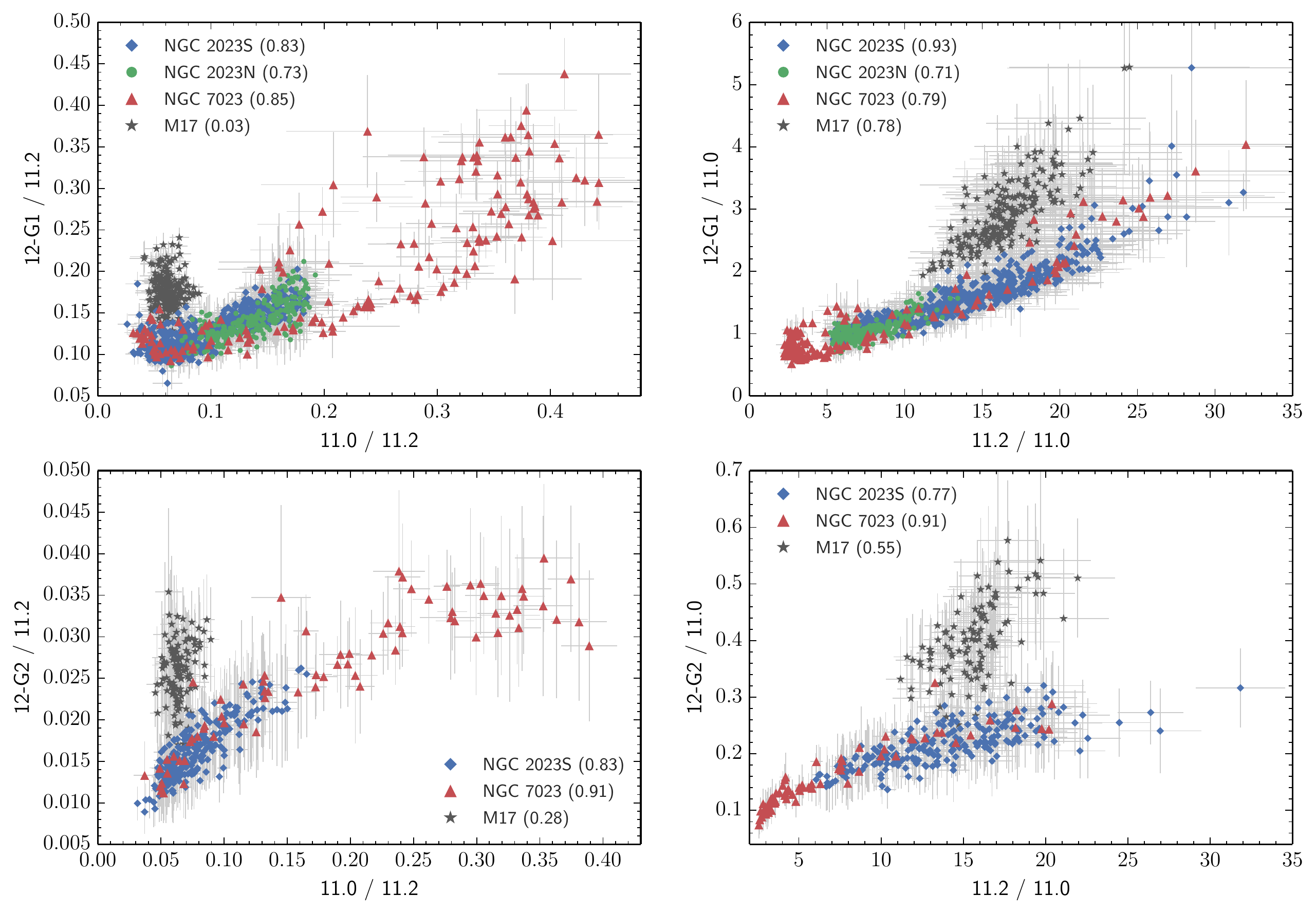}
    } \\

\caption{Correlation plots comparing the intensities of the 12-G1, 12-G2, 12-G3 and 12-G4 components of the 12.7 \mt decomposition to the traditional 11.0 and 11.2 \mt emission intensities. The number in parentheses is the weighted Pearson correlation coefficient for each source. Each figure is normalized to either the 11.0 or 11.2 \mt band, as appropriate.}
\label{fig:corr}
\end{center}
\end{figure*}

We examine ratios of band fluxes to probe for the presence of correlations between emission features (Fig.~\ref{fig:corr}). We compare first the 12-G3 and 12-G4 band fluxes to that of the 11.0 \mt emission (all normalized to the 11.2 \mt band). We also plot the 12-G3 and 12-G4 band fluxes against that of the 11.2 \mt band (then normalized to the 11.0 \mt emission). We identify a strong correlation between the 12-G4 and 11.2 \mt bands in NGC 7023 and NGC 2023 South and a mild correlation in NGC 2023 North (with associated weighted Pearson correlation coefficients of 0.98, 0.97 and 0.58, respectively). Another strong correlation is identified between the 12-G3 and 11.0 \mt bands in these same sources, with coefficients 0.96, 0.89 and 0.87, respectively. The ``cross terms" of these correlations, i.e. the 12-G3 band versus the 11.0 \mt band, and the 12-G3 versus the 11.2 \mt emission, reveal some residual structure and/or mild correlations. There is clearly some ``cross-contamination", but the strongest relationships are nevertheless between the 12-G3 and 11.0 \mt bands, and between the 12-G4 and the 11.2 \mt bands. Similar correlation plots for the 12-G1 and 12-G2 components are presented in the lower panels of Fig.~\ref{fig:corr}. In general, both the 12-G1 and 12-G2 components correlate with the 11.0 and 11.2 \mt bands, implying a mixed charge origin. M17 is an unusual case, as the only distinction between the traditional 11.0 and 11.2 \mt emission in the spectral maps was the M17 spur. The correlation plots involving M17 are also outlying cases, as the 11.0/11.2 ratio is constant.

\bibliographystyle{apj}

\begin{thebibliography}{49}
\expandafter\ifx\csname natexlab\endcsname\relax\def\natexlab#1{#1}\fi

\bibitem[{{Allamandola} {et~al.}(1999){Allamandola}, {Hudgins}, \&
  {Sandford}}]{allamandola1999}
{Allamandola}, L.~J., {Hudgins}, D.~M., \& {Sandford}, S.~A. 1999, \apjl, 511,
  L115

\bibitem[{{Allamandola} {et~al.}(1989){Allamandola}, {Tielens}, \&
  {Barker}}]{allamandola1989}
{Allamandola}, L.~J., {Tielens}, A.~G.~G.~M., \& {Barker}, J.~R. 1989, \apjs,
  71, 733

\bibitem[{{Bauschlicher} {et~al.}(2008){Bauschlicher}, {Peeters}, \&
  {Allamandola}}]{bauschlicher2008}
{Bauschlicher}, Jr., C.~W., {Peeters}, E., \& {Allamandola}, L.~J. 2008, \apj,
  678, 316

\bibitem[{{Bauschlicher} {et~al.}(2009){Bauschlicher}, {Peeters}, \&
  {Allamandola}}]{bauschlicher2009}
---. 2009, \apj, 697, 311

\bibitem[{{Bauschlicher} {et~al.}(2010){Bauschlicher}, {Boersma}, {Ricca},
  {Mattioda}, {Cami}, {Peeters}, {S{\'a}nchez de Armas}, {Puerta Saborido},
  {Hudgins}, \& {Allamandola}}]{bauschlicher2010}
{Bauschlicher}, Jr., C.~W., {et~al.} 2010, \apjs, 189, 341

\bibitem[{{Bern{\'e}} \& {Tielens}(2012)}]{berne2012}
{Bern{\'e}}, O., \& {Tielens}, A.~G.~G.~M. 2012, Proceedings of the National
  Academy of Science, 109, 401

\bibitem[{{Bern{\'e}} {et~al.}(2007){Bern{\'e}}, {Joblin}, {Deville}, {Smith},
  {Rapacioli}, {Bernard}, {Thomas}, {Reach}, \& {Abergel}}]{berne2007}
{Bern{\'e}}, O., {et~al.} 2007, \aap, 469, 575

\bibitem[{{Boersma} {et~al.}(2011){Boersma}, {Bauschlicher}, {Ricca},
  {Mattioda}, {Peeters}, {Tielens}, \& {Allamandola}}]{boersma2011}
{Boersma}, C., {Bauschlicher}, Jr., C.~W., {Ricca}, A., {et~al.} 2011, \apj,
  729, 64

\bibitem[{{Boersma} {et~al.}(2014{\natexlab{a}}){Boersma}, {Bregman}, \&
  {Allamandola}}]{boersma2014}
{Boersma}, C., {Bregman}, J., \& {Allamandola}, L.~J. 2014{\natexlab{a}}, \apj,
  795, 110

\bibitem[{{Boersma} {et~al.}(2015){Boersma}, {Bregman}, \&
  {Allamandola}}]{boersma2015}
---. 2015, \apj, 806, 121

\bibitem[{{Boersma} {et~al.}(2013){Boersma}, {Bregman}, \&
  {Allamandola}}]{boersma2013}
{Boersma}, C., {Bregman}, J.~D., \& {Allamandola}, L.~J. 2013, \apj, 769, 117

\bibitem[{{Boersma} {et~al.}(2012){Boersma}, {Rubin}, \&
  {Allamandola}}]{boersma2012}
{Boersma}, C., {Rubin}, R.~H., \& {Allamandola}, L.~J. 2012, \apj, 753, 168

\bibitem[{{Boersma} {et~al.}(2014{\natexlab{b}}){Boersma}, {Bauschlicher},
  {Ricca}, {Mattioda}, {Cami}, {Peeters}, {S{\'a}nchez de Armas}, {Puerta
  Saborido}, {Hudgins}, \& {Allamandola}}]{boersma2014_amesdb}
{Boersma}, C., {Bauschlicher}, Jr., C.~W., {Ricca}, A., {et~al.}
  2014{\natexlab{b}}, \apjs, 211, 8

\bibitem[{{Candian} {et~al.}(2012){Candian}, {Kerr}, {Song}, {McCombie}, \&
  {Sarre}}]{candian2012}
{Candian}, A., {Kerr}, T.~H., {Song}, I.-O., {McCombie}, J., \& {Sarre}, P.~J.
  2012, \mnras, 426, 389

\bibitem[{{Candian} \& {Sarre}(2015)}]{candian2015}
{Candian}, A., \& {Sarre}, P.~J. 2015, \mnras, 448, 2960

\bibitem[{{Chiar} \& {Tielens}(2006)}]{chiar2006}
{Chiar}, J.~E., \& {Tielens}, A.~G.~G.~M. 2006, \apj, 637, 774

\bibitem[{{Chini} {et~al.}(1980){Chini}, {Elsaesser}, \& {Neckel}}]{chini1980}
{Chini}, R., {Elsaesser}, H., \& {Neckel}, T. 1980, \aap, 91, 186

\bibitem[{{Galliano} {et~al.}(2008){Galliano}, {Madden}, {Tielens}, {Peeters},
  \& {Jones}}]{galliano2008b}
{Galliano}, F., {Madden}, S.~C., {Tielens}, A.~G.~G.~M., {Peeters}, E., \&
  {Jones}, A.~P. 2008, \apj, 679, 310

\bibitem[{{Hoffmeister} {et~al.}(2008){Hoffmeister}, {Chini}, {Scheyda},
  {Schulze}, {Watermann}, {N{\"u}rnberger}, \& {Vogt}}]{hoffmeister2008}
{Hoffmeister}, V.~H., {Chini}, R., {Scheyda}, C.~M., {Schulze}, D.,
  {Watermann}, R., {N{\"u}rnberger}, D., \& {Vogt}, N. 2008, \apj, 686, 310

\bibitem[{{Hony} {et~al.}(2001){Hony}, {Van Kerckhoven}, {Peeters}, {Tielens},
  {Hudgins}, \& {Allamandola}}]{hony2001}
{Hony}, S., {Van Kerckhoven}, C., {Peeters}, E., {Tielens}, A.~G.~G.~M.,
  {Hudgins}, D.~M., \& {Allamandola}, L.~J. 2001, \aap, 370, 1030

\bibitem[{{Houck} {et~al.}(2004){Houck}, {Roellig}, {van Cleve}, {Forrest},
  {Herter}, {Lawrence}, {Matthews}, {Reitsema}, {Soifer}, {Watson}, {Weedman},
  {Huisjen}, {Troeltzsch}, {Barry}, {Bernard-Salas}, {Blacken}, {Brandl},
  {Charmandaris}, {Devost}, {Gull}, {Hall}, {Henderson}, {Higdon}, {Pirger},
  {Schoenwald}, {Sloan}, {Uchida}, {Appleton}, {Armus}, {Burgdorf},
  {Fajardo-Acosta}, {Grillmair}, {Ingalls}, {Morris}, \& {Teplitz}}]{houck2004}
{Houck}, J.~R., {et~al.} 2004, \apjs, 154, 18

\bibitem[{{Hudgins} \& {Allamandola}(1999)}]{hudgins1999}
{Hudgins}, D.~M., \& {Allamandola}, L.~J. 1999, \apjl, 516, L41

\bibitem[{{Hudgins} {et~al.}(2005){Hudgins}, {Bauschlicher}, \&
  {Allamandola}}]{hudgins2005}
{Hudgins}, D.~M., {Bauschlicher}, Jr., C.~W., \& {Allamandola}, L.~J. 2005,
  \apj, 632, 316

\bibitem[{Hunter(2007)}]{hunter2007}
Hunter, J.~D. 2007, Computing In Science \& Engineering, 9, 90

\bibitem[{{Joalland} {et~al.}(2009){Joalland}, {Simon}, {Marsden}, \&
  {Joblin}}]{joalland2009}
{Joalland}, B., {Simon}, A., {Marsden}, C.~J., \& {Joblin}, C. 2009, \aap, 494,
  969

\bibitem[{{Joblin} {et~al.}(1996){Joblin}, {Tielens}, {Geballe}, \&
  {Wooden}}]{joblin1996}
{Joblin}, C., {Tielens}, A.~G.~G.~M., {Geballe}, T.~R., \& {Wooden}, D.~H.
  1996, \apjl, 460, L119

\bibitem[{{Jochims} {et~al.}(1994){Jochims}, {Ruhl}, {Baumgartel}, {Tobita}, \&
  {Leach}}]{jochims1994}
{Jochims}, H.~W., {Ruhl}, E., {Baumgartel}, H., {Tobita}, S., \& {Leach}, S.
  1994, \apj, 420, 307

\bibitem[{{Knorke} {et~al.}(2009){Knorke}, {Langer}, {Oomens}, \&
  {Dopfer}}]{knorke2009}
{Knorke}, H., {Langer}, J., {Oomens}, J., \& {Dopfer}, O. 2009, \apjl, 706, L66

\bibitem[{{Markwardt}(2009)}]{markwardt2009}
{Markwardt}, C.~B. 2009, in Astronomical Society of the Pacific Conference
  Series, Vol. 411, Astronomical Data Analysis Software and Systems XVIII, ed.
  D.~A. {Bohlender}, D.~{Durand}, \& P.~{Dowler}, 251

\bibitem[{{Mart{\'{\i}}n-Hern{\'a}ndez}
  {et~al.}(2002){Mart{\'{\i}}n-Hern{\'a}ndez}, {Peeters}, {Morisset},
  {Tielens}, {Cox}, {Roelfsema}, {Baluteau}, {Schaerer}, {Mathis}, {Damour},
  {Churchwell}, \& {Kessler}}]{martin-hernandez2002}
{Mart{\'{\i}}n-Hern{\'a}ndez}, N.~L., {et~al.} 2002, \aap, 381, 606

\bibitem[{Mor{\'e}(1978)}]{more1978}
Mor{\'e}, J.~J. 1978, in Lecture Notes in Mathematics, Vol. 630, Numerical
  Analysis, ed. G.~Watson (Springer Berlin Heidelberg), 105--116

\bibitem[{{Pech} {et~al.}(2002){Pech}, {Joblin}, \& {Boissel}}]{pech2002}
{Pech}, C., {Joblin}, C., \& {Boissel}, P. 2002, \aap, 388, 639

\bibitem[{{Peeters}(2011)}]{peeters2011}
{Peeters}, E. 2011, in IAU Symposium, Vol. 280, IAU Symposium, 149--161

\bibitem[{{Peeters} {et~al.}(2016){Peeters}, {Bauschlicher}, {Allamandola},
  {Tielens}, {Ricca}, \& {Wolfire}}]{peeters2016}
{Peeters}, E., {Bauschlicher}, Jr., C.~W., {Allamandola}, L.~J., {Tielens}, A.~G.~G.~M., {Ricca}, A., {Wolfire}, M.~G., 2016, \apj, Submitted

\bibitem[{{Peeters} {et~al.}(2002){Peeters}, {Hony}, {Van Kerckhoven},
  {Tielens}, {Allamandola}, {Hudgins}, \& {Bauschlicher}}]{peeters2002}
{Peeters}, E., {Hony}, S., {Van Kerckhoven}, C., {Tielens}, A.~G.~G.~M.,
  {Allamandola}, L.~J., {Hudgins}, D.~M., \& {Bauschlicher}, C.~W. 2002, \aap,
  390, 1089

\bibitem[{{Peeters} {et~al.}(2012){Peeters}, {Tielens}, {Allamandola}, \&
  {Wolfire}}]{peeters2012}
{Peeters}, E., {Tielens}, A.~G.~G.~M., {Allamandola}, L.~J., \& {Wolfire},
  M.~G. 2012, \apj, 747, 44

\bibitem[{{Povich} {et~al.}(2007){Povich}, {Stone}, {Churchwell}, {Zweibel},
  {Wolfire}, {Babler}, {Indebetouw}, {Meade}, \& {Whitney}}]{povich2007}
{Povich}, M.~S., {et~al.} 2007, \apj, 660, 346

\bibitem[{{Rapacioli} {et~al.}(2006){Rapacioli}, {Calvo}, {Joblin}, {Parneix},
  {Toublanc}, \& {Spiegelman}}]{rapacioli2006}
{Rapacioli}, M., {Calvo}, F., {Joblin}, C., {Parneix}, P., {Toublanc}, D., \&
  {Spiegelman}, F. 2006, \aap, 460, 519

\bibitem[{{Ricca} {et~al.}(2012){Ricca}, {Bauschlicher}, {Boersma}, {Tielens},
  \& {Allamandola}}]{ricca2012}
{Ricca}, A., {Bauschlicher}, Jr., C.~W., {Boersma}, C., {Tielens}, A.~G.~G.~M.,
  \& {Allamandola}, L.~J. 2012, \apj, 754, 75

\bibitem[{{Rosenberg} {et~al.}(2011){Rosenberg}, {Bern{\'e}}, {Boersma},
  {Allamandola}, \& {Tielens}}]{rosenberg2011}
{Rosenberg}, M.~J.~F., {Bern{\'e}}, O., {Boersma}, C., {Allamandola}, L.~J., \&
  {Tielens}, A.~G.~G.~M. 2011, \aap, 532, A128

\bibitem[{{Sellgren} {et~al.}(2007){Sellgren}, {Uchida}, \&
  {Werner}}]{sellgren2007}
{Sellgren}, K., {Uchida}, K.~I., \& {Werner}, M.~W. 2007, \apj, 659, 1338

\bibitem[{{Shannon} {et~al.}(2015){Shannon}, {Stock}, \&
  {Peeters}}]{shannon2015}
{Shannon}, M.~J., {Stock}, D.~J., \& {Peeters}, E. 2015, \apj, 811, 153

\bibitem[{{Sheffer} \& {Wolfire}(2013)}]{sheffer2013}
{Sheffer}, Y., \& {Wolfire}, M.~G. 2013, \apjl, 774, L14

\bibitem[{{Sloan} {et~al.}(1999){Sloan}, {Hayward}, {Allamandola}, {Bregman},
  {DeVito}, \& {Hudgins}}]{sloan1999}
{Sloan}, G.~C., {Hayward}, T.~L., {Allamandola}, L.~J., {et~al.} 1999, \apjl,
  513, L65

\bibitem[{{Sloan} {et~al.}(2007){Sloan}, {Jura}, {Duley}, {Kraemer},
  {Bernard-Salas}, {Forrest}, {Sargent}, {Li}, {Barry}, {Bohac}, {Watson}, \&
  {Houck}}]{sloan2007}
{Sloan}, G.~C., {et~al.} 2007, \apj, 664, 1144

\bibitem[{{Sloan} {et~al.}(2014){Sloan}, {Lagadec}, {Zijlstra}, {Kraemer},
  {Weis}, {Matsuura}, {Volk}, {Peeters}, {Duley}, {Cami}, {Bernard-Salas},
  {Kemper}, \& {Sahai}}]{sloan2014}
---. 2014, \apj, 791, 28

\bibitem[{{Smith} {et~al.}(2007{\natexlab{a}}){Smith}, {Armus}, {Dale},
  {Roussel}, {Sheth}, {Buckalew}, {Jarrett}, {Helou}, \&
  {Kennicutt}}]{smith2007cubism}
{Smith}, J.~D.~T., {Armus}, L., {Dale}, D.~A., {et~al.} 2007{\natexlab{a}},
  \pasp, 119, 1133

\bibitem[{{Smith} {et~al.}(2007{\natexlab{b}}){Smith}, {Draine}, {Dale},
  {Moustakas}, {Kennicutt}, {Helou}, {Armus}, {Roussel}, {Sheth}, {Bendo},
  {Buckalew}, {Calzetti}, {Engelbracht}, {Gordon}, {Hollenbach}, {Li},
  {Malhotra}, {Murphy}, \& {Walter}}]{smith2007}
{Smith}, J.~D.~T., {Draine}, B.~T., {Dale}, D.~A., {et~al.} 2007{\natexlab{b}},
  \apj, 656, 770

\bibitem[{{Spoon} {et~al.}(2007){Spoon}, {Marshall}, {Houck}, {Elitzur}, {Hao},
  {Armus}, {Brandl}, \& {Charmandaris}}]{spoon2007}
{Spoon}, H.~W.~W., {Marshall}, J.~A., {Houck}, J.~R., {et~al.} 2007, \apjl,
  654, L49

\bibitem[{{Stock} {et~al.}(2016){Stock}, {Choi}, {Moya}, {Otaguro}, {Sorkhou},
  {Allamandola}, {Tielens}, \& {Peeters}}]{stock2016}
{Stock}, D.~J., {Choi}, W.~D.-Y., {Moya}, L.~G.~V., {et~al.} 2016, \apj, 819,
  65  

\bibitem[{{Stock} {et~al.}(2014){Stock}, {Peeters}, {Choi}, \&
  {Shannon}}]{stock2014}
{Stock}, D.~J., {Peeters}, E., {Choi}, W.~D.-Y., \& {Shannon}, M.~J. 2014,
  \apj, 791, 99

\bibitem[{{Stock} {et~al.}(2013){Stock}, {Peeters}, {Tielens}, {Otaguro}, \&
  {Bik}}]{stock2013}
{Stock}, D.~J., {Peeters}, E., {Tielens}, A.~G.~G.~M., {Otaguro}, J.~N., \&
  {Bik}, A. 2013, \apj, 771, 72

\bibitem[{van~der Walt {et~al.}(2014)van~der Walt, {S}ch\"onberger,
  {Nunez-Iglesias}, {B}oulogne, {W}arner, {Y}ager, {G}ouillart, {Y}u, \& the
  scikit-image contributors}]{scikit-image}
van~der Walt, S., {S}ch\"onberger, J.~L., {Nunez-Iglesias}, J., {et~al.} 2014,
  PeerJ, 2, e453

\bibitem[{{van der Zwet} \& {Allamandola}(1985)}]{vanderzwet1985}
{van der Zwet}, G.~P., \& {Allamandola}, L.~J. 1985, \aap, 146, 76

\bibitem[{{van Diedenhoven} {et~al.}(2004){van Diedenhoven}, {Peeters}, {Van
  Kerckhoven}, {Hony}, {Hudgins}, {Allamandola}, \&
  {Tielens}}]{vandiedenhoven2004}
{van Diedenhoven}, B., {Peeters}, E., {Van Kerckhoven}, C., {Hony}, S.,
  {Hudgins}, D.~M., {Allamandola}, L.~J., \& {Tielens}, A.~G.~G.~M. 2004, \apj,
  611, 928

\bibitem[{{Van Kerckhoven} {et~al.}(2000){Van Kerckhoven}, {Hony}, {Peeters},
  {Tielens}, {Allamandola}, {Hudgins}, {Cox}, {Roelfsema}, {Voors}, {Waelkens},
  {Waters}, \& {Wesselius}}]{vankerckhoven2000}
{Van Kerckhoven}, C., {et~al.} 2000, \aap, 357, 1013

\bibitem[{{Wang} {et~al.}(2004){Wang}, {Bovik}, {Sheikh}, \&
  {Simoncelli}}]{wang2004}
{Wang}, Z., {Bovik}, A.~C., {Sheikh}, H.~R., \& {Simoncelli}, E.~P. 2004, IEEE
  Transactions on Image Processing, 13, 600

\bibitem[{{Werner} {et~al.}(2004){Werner}, {Roellig}, {Low}, {Rieke}, {Rieke},
  {Hoffmann}, {Young}, {Houck}, {Brandl}, {Fazio}, {Hora}, {Gehrz}, {Helou},
  {Soifer}, {Stauffer}, {Keene}, {Eisenhardt}, {Gallagher}, {Gautier}, {Irace},
  {Lawrence}, {Simmons}, {Van Cleve}, {Jura}, {Wright}, \&
  {Cruikshank}}]{werner2004}
{Werner}, M.~W., {et~al.} 2004, \apjs, 154, 1

\bibitem[{{Xu} {et~al.}(2011){Xu}, {Moscadelli}, {Reid}, {Menten}, {Zhang},
  {Zheng}, \& {Brunthaler}}]{xu2011}
{Xu}, Y., {Moscadelli}, L., {Reid}, M.~J., {Menten}, K.~M., {Zhang}, B.,
  {Zheng}, X.~W., \& {Brunthaler}, A. 2011, \apj, 733, 25

\end{thebibliography}

\end{document}